\documentclass[firstname=Dhruv\ Bhatia-Kara \ Fagerstrom-Max\  Watson, lastname=, course=Rose\  Hulman\  REU, showname=True, showemail=False, brownid=dbhatia1, num=, type=Report]{homework}

\usepackage{kbordermatrix}
\usepackage{xcolor}
\usepackage{tikz}
\usepackage[backend=biber,style=alphabetic,sorting=nty]{biblatex}
\usepackage{verbatim}
\usepackage{cleveref}

\usepackage{hw-shortcuts}
\title{Probability Distributions for Elliptic Curves in the CGL Hash Function}
\author{Dhruv Bhatia, Kara Fagerstrom, Max Watson \\ \lineskip 1em \textbf{Advisor}: Joshua Holden}

\titleformat{\section}{\Large\sc}{\thesection}{1em}{}
\titlespacing{\section}{0em}{0em}{1em}

\titleformat{\subsection}{\large\bf}{\thesubsection}{1em}{}
\titlespacing{\subsection}{0em}{0em}{1em}

\begin{document}
\maketitle
\begin{abstract}
    Hash functions map data of arbitrary length to data of predetermined length. Good hash functions are hard to predict, making them useful in cryptography. We are interested in the elliptic curve CGL hash function, which maps a bitstring to an elliptic curve by traversing an input-determined path through an isogeny graph. The nodes of an isogeny graph are elliptic curves, and the edges are special maps betwixt elliptic curves called isogenies. Knowing which hash values are most likely informs us of potential security weaknesses in the hash function. We use stochastic matrices to compute the expected probability distributions of the hash values. We generalize our experimental data into a theorem that completely describes all possible probability distributions of the CGL hash function. We use this theorem to evaluate the collision resistance of the CGL hash function and compare this to the collision resistance of an ``ideal" hash function.
\end{abstract}
\pagebreak
\tableofcontents
\section{Introduction}\label{Sec:introduction}
    Hash functions are a way of mapping arbitrarily long data to data of a predetermined length in a way that preserves uniqueness. The idea is that small changes in the input should result in much more drastic changes in the output. Functions like these are extremely useful and have many applications in computer science and cryptography. 
    
    \pagebreak
    
    For example, in computer science, hash functions are used to quickly store and access data by mapping data to a memory address. If we want to store some information, we can compute the hash value of the data and store the data at that memory address. Later, to look up this data, instead of serially searching through all the memory addresses, we can simply compute the hash value again. Thus, a good hash function is quick to compute and has a low chance of two random pieces of data colliding at the same hash value.
    
    Hash functions can also be used to commit to data without revealing it. For example, if two parties are bidding for the same item, it would be nice if both parties could place bids without revealing the amounts of the bids. This way the parties do not influence each other in any way. Here, both parties could place their bids and only reveal the hash values of the bid. Later, when the bids are revealed, the hash values can be recomputed and checked against the original values, ensuring that the bids weren't altered at any time. So, a good hash function is difficult to reverse, and it should also be difficult to ``engineer" data that has a particular hash value.
    
    In \cite{CGL}, Charles, Goren, and Lauter created a hash function that maps data to a finite set of elliptic curves by computing special maps called isogenies between elliptic curves.  In \cref{Sec:background}, we provide background on elliptic curves, isogenies, and the mechanics of the CGL hash function.
    
    At a high level, the CGL hash function works by following a series of maps between elliptic curves. To better study the hash function, we can create graphs, called isogeny graphs, illustrating all possible maps. In \cref{Sec:graphs}, we outline our algorithm for creating these graphs. This algorithm has been implemented in SageMath at \href{https://github.com/dhruvbhatia00/CGL-Hash.git}{https://github.com/dhruvbhatia00/CGL-Hash.git}.
    
    To evaluate its security, we analyze how difficult the CGL hash function is to predict. In particular, we wish to find a probability distribution describing how likely it is for a random input to have a particular hash value. In \cref{Sec:stochastic}, we describe a method of computing these probability distributions using stochastic matrices. Next, we generalize our computational results into a theorem about these probability distributions, which we prove in \cref{Sec:distributions}. Finally, in \cref{Sec:conclusions}, we discuss the implications of our theorem on the collision resistance of the CGL hash function and outline potential directions of future work.
    
\section{Background} \label{Sec:background}
    We begin in \cref{Subsec:curves} with background on elliptic curves and the maps between them called isogenies. Then, in \cref{Subsec:velu}, we explain V\'{e}lu's formulae for computing isogenies. In \cref{Subsec:dual}, we define the dual of an isogeny and show some of its properties. Finally, in \cref{Subsec:CGL}, we explain the algorithm used in the CGL hash function.
    
    \pagebreak
    
    \subsection{Elliptic Curves and Isogenies}\label{Subsec:curves}
    Elliptic curves are a special type of curve living in the plane. Elliptic curves can be described by a class of equations called Weierstrass equations. However, not all Weierstrass equations describe elliptic curves, as elliptic curves come with a few extra restrictions and properties.
        \begin{Def}
            A \textbf{Weierstrass equation} defined over a field $K$ is an equation of the form $Y^2 + a_1XY + a_3Y = X^3 + a_2X^2 + a_4X + a_6$, where $a_1, a_2, a_3, a_4, a_6 \in K$.
        \end{Def}
        
        Such equations describe a broad range of curves in the plane. In order to restrict ourselves to elliptic curves, we only look at those curves that have no cusps or self-intersections. To do this, we look at the discriminant of such equations.
        
        \begin{Def}\cite[Sec.~III.1]{Silverman}
            The \textbf{discriminant} of a Weierstrass equation $E: Y^2 + a_1XY + a_3Y = X^3 + a_2X^2 + a_4X + a_6$ is
                $$\Delta(E) = -b_2^2b_8 - 8b_4^3 - 27b_6^2 + 9b_2b_4b_6$$
            where 
            \begin{multicols}{2}
                \noindent\begin{align*}
                    b_2 &= a_1^2 + 4a_4 \\
                    b_4 &= 2a_4 + a_1a_3
                \end{align*}    
                \begin{align*}
                    b_6 &=a_3^2 + 4a_6 \\ 
                    b_8 &=a_1^2a_6+4a_2a_6-a_1a_3a_4+a_2a_3^2-a_4^2
                \end{align*}
            \end{multicols}
        \end{Def} 
        
        \begin{Def} \cite[Sec.~III.1]{Silverman}
            An \textbf{elliptic curve} defined over a field $K$ is a collection of points $(X,Y) \in K^2$ satisfying a Weierstrass equation defined over $K$ such that the discriminant is non-zero. Elliptic curves also contain an additional point ``at infinity'', denoted $\mathcal{O}_E$.
        \end{Def} 
        
        \begin{DefProp}\cite[Sec.~III.1]{Silverman}
            If $\char{K} \neq 2,3$ then a change of variables allows us to rewrite the equation in the form $y^2 = x^3 + ax + b$. This is called the \textbf{normal} form of an elliptic curve, and cubic polynomial on the right side of the equation is said to be a \textbf{depressed} cubic as it lacks an $x^2$ term.
        \end{DefProp} 
        
        In this form, the discriminant becomes:
                $$\Delta(E) = -16 \lrp{4a^3 + 27b^2}$$
        We see that this is the same formula as the discriminant for a depressed cubic, so requiring that an elliptic curve has non-zero discriminant is the same as asking that the cubic $x^3 + ax + b$ has no repeated roots.
        
        In this paper, we will only be dealing with curves defined over fields of characteristic $\neq 2,3$, so we can restrict to curves written in normal form.
        
        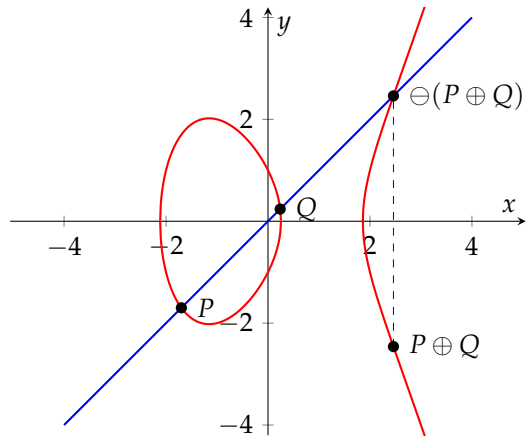
\begin{figure}[h]
            \centering
            \begin{tikzpicture}
                \begin{axis}[
                    xmin=-4.2,
                    xmax=4.2,
                    ymin=-4.2,
                    ymax=4.2,
                    axis equal,
                    axis lines=middle,
                    xlabel=$x$,
                    ylabel=$y$,
                    disabledatascaling,
                    trig format plots=rad,]
                       
                        \addplot[domain=1.86080601:4, samples = 300, red, thick]{sqrt(x^3 - 4*x + 1)};
                        \addplot[domain=1.86080601:4, samples = 300, red, thick]{-sqrt(x^3 - 4*x + 1)};
                        \addplot[domain=-2.1149:.2541016883, samples = 200, red, thick]{sqrt(x^3 - 4*x + 1)};
                        \addplot[domain=-2.1149:.2541016883, samples = 200, red, thick]{-sqrt(x^3 - 4*x + 1)};
                        
                        \addplot[domain=-4:4, samples = 200, blue, thick]{x};
                        \addplot[domain=-2.461:2.461, samples = 200, blue, dashed]({2.461}, {x});
                        \node[label={0:{$P$}},circle,fill,inner sep=1.5pt] at (axis cs:-1.699,-1.699) {};
                        \node[label={0:{$Q$}},circle,fill,inner sep=1.5pt] at (axis cs:0.2391,0.2391) {};
                        \node[label={0:{$\ominus (P \oplus Q)$}},circle,fill,inner sep=1.5pt] at (axis cs:2.4605,2.4605) {};
                        \node[label={0:{$P \oplus Q$}},circle,fill,inner sep=1.5pt] at (axis cs:2.4605,-2.4605) {};
                    \end{axis}
            \end{tikzpicture}
            \caption{Graph of $E: y^2 = x^3 - 4x + 1$ over $\R$}
            \label{fig:group_law}
        \end{figure}
        \pagebreak
        
        Given an elliptic curve $E$ over $\R$, we can define a group operation $\oplus$ on the points of the curve by setting the sum of any three co-linear points to be $\mathcal{O}_E$. In this way, $\mathcal{O}_E$ becomes the identity element of the group. More concretely, to add points $P$ and $Q$ on the curve $E$, we first find the line through them and find where this line intersects the curve a third time. We then reflect the third point about the $x$-axis to obtain $P \oplus Q$. This is illustrated in \cref{fig:group_law}. In order to add $P$ to itself, we would use the tangent line to $E$ at $P$. Finally, we see that all vertical lines through the curve intersect $E$ at at most two points in $\R^2$, and so we say that such lines also intersect the curve at $\mathcal{O}_E$. We conclude that $P$ and $Q$ will be inverse to one another if and only if the line through them is vertical. 
        
        We can write down formulae to describe this group law. Let $P = (x_P, y_p)$, $Q = (x_Q, y_Q)$ be points on $E: y^2 = x^3 + ax + b$. To start, if $x_P = x_Q$ and $y_P = -y_Q$, implying that $P$ and $Q$ are reflections of one another about the $x$-axis, then we set $P \oplus Q = \mathcal{O}_E$. We could also write $P = \ominus Q$ to mean that $Q$ is the inverse of $P$. Otherwise, we define a value $s$ as follows:
            \begin{equation*}
                s = \begin{cases}
                    \dfrac{y_P - y_Q}{x_P - x_Q} & \text{if $P \neq Q$} \\
                    \dfrac{3x_P^2 + a}{2y_P} & \text{if $P = Q$}
                \end{cases}
            \end{equation*}
        where $s$ describes the slope of the line between $P$ and $Q$. We then set $P \oplus Q = R = (x_R, y_R)$, where
            \begin{align*}
                x_R &= s^2 - x_P - x_Q \\
                y_R &= y_P + s(x_R - x_P)
            \end{align*}
        These formulae induce a group structure on the elliptic curve, irrespective of which field the curve is defined over. \cite[Sec.~III.2]{Silverman}
        
        We now define the structure preserving maps between elliptic curves. But what structure are we interested in preserving? Elliptic curves are described by polynomial equations, and so we might ask that the maps between them can be written as polynomials, or as rational functions. More importantly, elliptic curves are groups, and so we might ask that our maps are group homomorphisms. As we will see, the following definition encompasses both these ideas.
        
        \begin{Def}\cite[Sec.~III.4]{Silverman}
            An \textbf{isogeny} between two elliptic curves $E,E'$ defined over a field $K$ is a function $\phi: E \to E'$ given by
                \begin{equation*}
                    (x,y) \mapsto (p(x,y), q(x,y))
                \end{equation*}
            where $p$ and $q$ are rational functions over $K$ such that $\phi(\mathcal{O}_E) = \mathcal{O}_{E'}$. 
        \end{Def}
        
        \begin{Prop}\cite[Sec.~III.4.8]{Silverman}
            An isogeny $\phi: E \to E'$, with $E, E'$ defined over $K$ is a group homomorphism with finite kernel. When viewed over $\bar{K}$, this homomorphism is surjective.
        \end{Prop}
        
        \begin{Def}\cite[Sec.~9.3]{Galbraith}
            An \textbf{isomorphism} $\psi$ between elliptic curves $E$ and $E'$ is an invertible isogeny. That is, there exists an isogeny $\psi^{-1}: E \to E'$ to $E$ such that for all $P\in E$, $\psi^{-1}(\psi(P))=P$ and for all $Q \in E'$, $\psi(\psi^{-1}(Q)) = Q$. 
        \end{Def}
        
        An isomorphism is nothing more than a change of variables, so two isomorphic curves can be thought of as being ``the same." But how can we tell when two curves are isomorphic? The following gives us a quick, computational method of checking when two curves are isomorphic.
        
        \begin{Def} \cite[Sec.~III.1]{Silverman}
            The \textbf{$j$-invariant} of an elliptic curve $E:y^2=x^3+ax+b$ is defined by the equation: \begin{equation*}
                j(E) = 1728\frac{4a^3}{4a^3+27b^2}
            \end{equation*} 
        \end{Def}
        
        \begin{Prop}\cite[Sec.~III.1.4]{Silverman}
            Two elliptic curves defined over a field $K$ are isomorphic over the algebraic closure $\Bar{K}$ if and only if their $j$-invariants are the same.  
        \end{Prop}
        
        \begin{Def}
            Isogenies from a curve $E$ to itself are called \textbf{endomorphisms}. If an endomorphism is also an isomorphism, it is called an \textbf{automorphism}.
        \end{Def}

        \begin{Prop}\cite[Sec.~2.2]{Adj}
            Every isogeny $\phi: E \to E'$, with $E, E'$ elliptic curves over $K$ can be written in the form
                $$\phi(x,y) = \lrp{\frac{p_1(x)}{p_2(x)}, y\frac{q_1(x)}{q_2(x)}}$$
            where $p_1,p_2,q_1,q_2 \in K[x]$.
        \end{Prop}

        \begin{Def} \cite[Sec.~2.2]{Adj}
            Given an isogeny $\phi: E \to E'$ of form
                $$\phi(x,y) = \lrp{\frac{p_1(x)}{p_2(x)}, y\frac{q_1(x)}{q_2(x)}}$$
            the \textbf{degree} of the isogeny is $\deg\lrp{\phi} = \max\lrp{\deg(p_1), \deg(p_2)}$. An isogeny is called \textbf{separable} if
                $$\frac{d}{dx} \frac{p_1(x)}{p_2(x)} \neq 0$$
            Otherwise, the isogeny is called \textbf{inseparable}.
        \end{Def}
        
        \begin{Prop}\cite[Sec.~III.4]{Silverman}
            Let $\phi: E \to E'$ and $\psi: E' \to E''$ be isogenies. Then $\deg\lrp{\psi \circ \phi} = \deg\lrp{\psi} \cdot \deg\lrp{\phi}$.
        \end{Prop}
        
        \begin{Prop}\cite[Sec.~25.1]{Galbraith}
            Let $\phi: E \to E'$ be an isogeny. Then $|\ker{\phi}|$ divides $\deg(\phi)$. If $\phi$ is separable, then $|\ker{\phi}| = \deg(\phi)$.
        \end{Prop}
        
        As we will see in \cref{Subsec:CGL}, isogenies are the building blocks of the hash function described in \cite{CGL}. The following section describes how, given an elliptic curve $E$, we can easily compute the separable isogenies out of $E$ (this is especially easy when we only care about degree $2$ isogenies). However, given two elliptic curves $E_1$ and $E_2$, it is much harder to tell whether there exists an isogeny between them, and even harder still to compute the isogeny if it exists \cite[Sec.~5.3]{CGL}. It is this property of isogenies that makes the hash function quick and easy to compute, but very difficult to reverse.

    \subsection{V\'{e}lu's Formulae}\label{Subsec:velu}
        Every isogeny, being a group homomorphism, has a kernel. But can we go backwards? Can we start with a subgroup $G$ of a curve and find an isogeny out of that curve with kernel $G$?
        
        \begin{Prop}\cite[Sec.~III.4.12]{Silverman}
            Given a finite subgroup $G$ of an elliptic curve $E$, there is an elliptic curve $E'$ (unique up to isomorphism), along with a separable isogeny $\phi: E \to E'$ with kernel $G$ (unique up to post-composition by the same isomorphism).
        \end{Prop}

        The above proposition implies that isogenies are uniquely defined (up to isomorphism) by their kernels. V\'{e}lu's formulae give us a way of taking a finite subgroup $G$ of an elliptic curve $E: y^2 = f(x)$, and explicitly computing an elliptic curve $E'$, along with a separable isogeny $\phi: E \to E'$ such that $\phi$ has kernel $G$. In this paper, we will be looking at isogenies of degree 2, in which the kernel $G$ contains the identity $\mathcal{O}_E$ and an order 2 point on the same elliptic curve. We restrict ourselves to points of order $2$ because they are easy to compute - they all have the form $(x_0, 0)$, where $x_0$ is a root of $f(x)$.
        
        Let $E: y^2 = f(x) = x^3 + ax + b$ be an elliptic curve defined over a field $K$. Viewing the curve over the field $\Bar{K}$, let $P = (x_P, y_P)$ be an order $2$ point. Since $P$ has order $2$, $P \oplus P = \mathcal{O}_E$, implying that the tangent line to $E$ at $P$ is vertical. But by the vertical symmetry of the curve, this can only happen when $y_P = 0$. So, $P = (x_P, 0)$, where $x_P$ is a root of $f(x)$.
        
        The formulae presented below have been adjusted to reflect the specific form of kernel we are interested in, but the originals can be found in \cite[Sec.~25.1.1]{Galbraith}.
        
        We can define a new elliptic curve $E': Y^2 = X^3 + AX + B$, where 
            \begin{align*}
                A &= -15x_P^2 - 4a \\
                B &= 8b - 14x_0^3
            \end{align*}
        V\'{e}lu also supplies us with the required isogeny between them defined as 
            \begin{align*}
                (x, y) &\mapsto \lrp{x + \frac{3x_P^2 + a}{x - x_P}, y - \frac{y(3x_P^2 + a)}{(x - x_p)^2}}
            \end{align*}
        The proof that this is in fact a separable isogeny between $E$ and $E'$ with kernel $\lrc{\mathcal{O}_E, P}$ can be found in \cite[Sec.~25.1.6]{Galbraith}
        
        \begin{Ex}
            Consider the curve $E: y^2 = f(x) = x^3 - 4x$ defined over $\R$. We see that $f(x)$ has a root $-2$, which we can plug into V\'{e}lu's formulae to obtain a new curve $\Tilde{E}: y^2 = x^3 + Ax + B$ where
                \begin{align*}
                    A &= -15 \cdot (-2)^2 - 4 \cdot (-4) = -44 \\
                    B &= 8 \cdot 0 - 14 \cdot (-2)^3 = 112
                \end{align*}
            Then, $E: y^2 = x^3 - 4x$ and $\Tilde{E}: y^2 = x^3 - 44x + 112$ have a degree $2$ isogeny $\phi: E \to \Tilde{E}$ given by
                $$(x,y) \mapsto \lrp{x + \frac{8}{x + 2}, y - \frac{8y}{(x+2)^2}}$$
            To see this isogeny in action, click here: \href{https://www.desmos.com/calculator/1eowvib3ov}{https://www.desmos.com/calculator/1eowvib3ov}. Here, the red curve is $E$, and the blue curve is $E'$. After choosing points $P$ and $Q$ on the red curve, we can see how they add to $P \oplus Q$ using the group law. The graph shows where the isogeny $\phi$ takes these three points, and we can see that indeed, $\phi(P \oplus Q) = \phi(P) \oplus \phi(Q)$.
        \end{Ex}
        
    \subsection{Dual Isogenies}\label{Subsec:dual}
        Before we introduce the CGL hash function, we need to talk about one more property of isogenies: for every isogeny $\phi: E \to E'$, there is another isogeny $\psi: E' \to E$, called the dual, such that the composition $\psi \circ \phi$ is given by $P \mapsto d \cdot P$, where $d = \deg{\phi}$. This is a rather surprising fact, and it allows us to think of the degree of an isogeny as a measure of how far the isogeny is from being an isomorphism --- degree $1$ isogenies are isomorphisms because after composing by the dual, we get the identity map. 
        
        \begin{Prop}\cite[Sec.~III.4.1]{Silverman}
            Given an elliptic curve $E$ and an integer $m$, the map $[m]: E \to E$ given by 
                $$P \mapsto \begin{cases}
                    m \cdot P & \text{when } m \geq 0 \\
                    -m \cdot (\ominus P) & \text{when } m < 0
                \end{cases}$$
            (here multiplication refers to repeated elliptic curve addition) is an isogeny of degree $m^2$.
        \end{Prop}
        
        \begin{Def}\cite[Sec.~9.1]{Galbraith}
            The \textbf{$m$-torsion subgroup} $E[m]$ of an elliptic curve $E$ over a field $K$ is the group of all points $P$ on $E$ such that $m \cdot P=\mathcal{O}_E$. Each such point $P$ is called an \textbf{$m$-torsion point} of $E$.
        \end{Def} 
        
        We can see that the kernel of the map $[m]$ is exactly $E[m]$, as both contain exactly those points sent to the identity after being multiplied by $m$.

        \begin{DefProp}\cite[Sec.~III.6.1]{Silverman}
            Every isogeny $\phi: E \to E'$ has a unique (up to post-composition by an automorphism) \textbf{dual isogeny} $\hat{\phi}: E' \to E$ with $\deg(\phi) = \deg(\hat{\phi})$ such that $\phi \circ \hat{\phi} = [\deg\lrp{\phi}]_E$ and $\hat{\phi} \circ \phi = [\deg\lrp{\phi}]_{E'}$.
        \end{DefProp}

        \pagebreak
        \begin{Prop}
            Let $E: y^2 = f(x)$ be an elliptic curve, and let $x_1, x_2, x_3$ be the roots of $f(x)$. Let $\phi_1: E \to E_1$ be the isogeny out of $E$ with kernel $\lrc{\mathcal{O}_E, (x_1, 0)}$. Then $\phi(x_2, 0) = \phi(x_3, 0)$, and $\phi(x_2, 0)$ is a $2$-torsion point of $E_1$. Further, let $\psi: E_1 \to E_2$ be the isogeny out of $E_1$ with kernel $\lrc{\mathcal{O}_{E_1}, \phi(x_2, 0)}$. Then, $E_2$ is isomorphic to $E$ and $\psi$ is the dual of $\phi_1$ (up to composition by the isomorphism).
        \end{Prop}
        
        
        \begin{proof}
            Let $P_i = (x_i, 0)$. Then, the subgroup of $2$-torsion elements is $\lrc{\mathcal{O}_E, P_1, P_2, P_3}$. Since each $P_i$ has order $2$, this subgroup is isomorphic to the Klein four-group, and so adding any two non-zero points in the group yields the third one. 
            
            Since the kernel of $\phi_1$ is $\lrc{\mathcal{O}_E, P}$, we see that $\phi_1(P_1) = \mathcal{O}_{E_1}$. But
                $$\phi_1(P_3) = \phi_1(P_1) \oplus \phi_1(P_2) = \mathcal{O}_{E_1} \oplus \phi_1(P_2) = \phi_1(P_2)$$
            This proves the first claim. Next, we can see that
                $$2 \cdot \phi_1(P_2) = \phi_1(2 \cdot P_2) = \phi_1(\mathcal{O}_E) = \mathcal{O}_{E_1}$$
            showing that $\phi_1(P_2) = \phi_1(P_3)$ is a $2$-torsion point of $E_1$. Now, let $\psi$ be as defined above. We must show that $\psi \circ \phi_1 = [2]_E$. That $\phi_1 \circ \psi = [2]_{E_1}$ will follow a symmetrical argument. Since isogenies are uniquely defined by their kernels, it suffices to show that $\ker{\psi \circ \phi} = \ker{[2]_E}$.
            
            We know that the kernel of $[2]_E$ is the set of all points such that doubling the point turns it into the identity. In other words, $\ker{[2]_E}$ is the group of $2$-torsion points $\lrc{\mathcal{O}_E, P_1, P_2, P_3}$. Working case-by-case:
                \begin{align*}
                    \psi \circ \phi_1(\mathcal{O}_E) &= \mathcal{O}_E \\
                    \psi \circ \phi_1(P_1) &= \psi(\mathcal{O}_{E_1}) = \mathcal{O}_E \\
                    \psi \circ \phi_1(P_2) &= \mathcal{O}_E \\
                    \psi \circ \phi_1(P_3) &= \mathcal{O}_E 
                \end{align*}
            where the last two equations follow from the fact that $\phi_1(P_2) = \phi_1(P_3) \in \ker{\psi}.$ Therefore, $\ker{[2]_E} \subseteq \ker{\psi \circ \phi_1}$. 
            
            Next, let $P \in \ker{\psi \circ \phi_1}$. Then, $\psi \circ \phi_1(P) = \mathcal{O}_{E_2}$, and so $\phi_1(P) \in \ker{\psi} = \lrc{\mathcal{O}_{E_1},\phi_1(P_2)}$. We work in cases:
                \begin{itemize}
                    \item If $\phi_1(P) = \mathcal{O}_{E_1}$, then $P \in \ker{\phi_1} = \lrc{\mathcal{O}_E, P_1} \subseteq \ker{[2]_E}$.
                    \item If $\phi_1(P) = \phi_1(P_2)$, then $\phi_1(P - P_2) = \mathcal{O}_{E_1}$, implying that $P - P_2 \in \ker{\phi_1} = \lrc{\mathcal{O}_E, P_1}$, and so $P \in \lrc{P_2, P_3} \subseteq \ker{[2]_E}$.
                \end{itemize}
            Therefore, $\ker{\psi \circ \phi_1} \subseteq \ker{[2]_E}$. We conclude that $\psi$ is indeed the dual of $\phi$. Since duals are unique up to post-composition by an isomorphism, $E_2$ must be isomorphic to $E$. This completes the proof.
        \end{proof}
        
        \begin{Ex}
            Earlier, we saw the example of the curve $E: y^2 = f(x) = x^3 - 4x$ defined over $\R$. We used the root $-2$ to create the isogeny $\phi: E \to \Tilde{E}$, where $\Tilde{E}: y^2 = x^3 - 44x + 112$, and $\phi$ is given by:
                $$(x,y) \mapsto \lrp{x + \frac{8}{x + 2}, y - \frac{8y}{(x+2)^2}}$$
            By the above proposition, we should be able to compute the dual $\hat{\phi}$ as the isogeny out of $\Tilde{E}$ with kernel $\lrc{\mathcal{O}_{\Tilde{E}}, \phi(x_2, 0)}$, where $x_2 \neq -2$ is another root of $f(x)$. In this example, we see that $0$ is another root, and so $\phi(0,0) = (4, 0)$. Since $(4,0)$ has $0$ in the $y$-coordinate, it is a $2$-torsion point of $\Tilde{E}$, as described in the proposition. So, we can plug $4$ into V\'{e}lu's formulae, this time to go in the other direction. 
            
            V\'{e}lu's formulae give us a new curve $E': y^2 = x^3 - 64x$ and a map $\psi: \Tilde{E} \to E'$. Since $E'$ and $E$ both have $j$-invariant $0$, they are isomorphic. Post composing $\psi$ with this isomorphism yields $\hat{\phi}: \Tilde{E} \to E$ given by
                $$(x,y) \mapsto \lrp{\dfrac{\frac{1}{4}x^2 - x + 1}{x - 4}, y \cdot \dfrac{\frac{1}{8}x^2 - x + \frac{3}{2}}{x^2 - 8x + 16}}$$
            Consider the composition $\hat{\phi} \circ \phi$. We see that the $\ker{\hat{\phi} \circ \phi} = \lrc{(x,y) \in E : \phi(x,y) \in \ker{\hat{\phi}}}$.
            
            By definition, $\ker{\hat{\phi}} = \lrc{\mathcal{O}_{\Tilde{E}}, (4,0)}$. So, to compute $\ker{\hat{\phi} \circ \phi}$, we need to find points $(x,y)$ of $E$ with an $x$-coordinate of $4$ after being hit by $\phi$:
                \begin{align*}
                    x + \frac{8}{x + 2} &= 4 \\
                    \frac{x^2 + 2x + 8}{x + 2} &= 4 \\
                    x^2 + 2x + 8 &= 4x + 8 \\
                    x(x - 2) &= 0
                \end{align*}
            Such points are those with $x = 0$ or $x = 2$. Plugging these into the equation describing $E$, we see that the points are $(0,0)$ and $(2, 0)$. Finally, we also note that $\phi(-2, 0) = \mathcal{O}_{\Tilde{E}}$, and so $\ker{\hat{\phi} \circ \phi} = \lrc{\mathcal{O}_E, (0,0), (2,0), (-2, 0)}$, which is exactly the $2$-torsion subgroup of $E$, showing that $\hat{\phi} \circ \phi = [2]$.
        \end{Ex}
    
    \subsection{CGL Hash Function}\label{Subsec:CGL}
    
        \begin{Def}
            A \textbf{hash function} is a function $f: B \to X$, where $B$ is the set of finitely long bitstrings, and $X$ is any finite set. 
        \end{Def}
        
        \pagebreak
        The idea is to have a way of taking data of arbitrary length and associating to it a value of fixed size in a way that preserves uniqueness. As discussed in \cref{Sec:introduction}, a good hash function $f$ has the following properties: \cite[Sec.~9.2.2]{hac}
            \begin{itemize}
                \item Hash values should be quick to compute.
                \item Given a randomly chosen bitstring, the likelihood of attaining a certain hash value should be evenly distributed among all hash values.
                \item Pre-image resistance: Given a has value $y$, it should be hard to find a bitstring $b$ such that $f(b) = y$.
                \item Second pre-image resistance: Given a bitstring $b_1$, it should be hard to find a second bitstring $b_2$ with $f(b_1) = f(b_2)$.
                \item Collision resistance: Given no starting information, it should be hard to find two bitstrings $b_1$ and $b_2$ such that $f(b_1) = f(b_2)$.
            \end{itemize}
        
        In \cite{CGL}, Charles, Goren and Lauter came up with a hash function which we will refer to as the CGL hash. To define the function, we must first choose a field $K$, along with an elliptic curve $E: y^2 = f(x)$, called the initial node, defined over it. We also order the three roots of $f(x)$ (which exist in some extension of $K$) and choose the first root $x_1$. Given a bitstring $b$ (a string of $1$s and $0$s), the function repeats the following for each bit in $b$: 
            \begin{enumerate}
                \item Let $x$ equal $x_2$ if the current bit is $0$ and $x_3$ if the current bit is $1$.
                \item Using V\'{e}lu's formulae, use $G=\{\mathcal{O},(x,0)\}$ to find an isogeny $\phi$ from $E$ to $E'$.
                \item Let $x_1=\phi(x_1)$, $E=E'$, let $x_2,x_3$ be the remaining two roots of $E'$ and repeat using the next bit of $b$.
            \end{enumerate}
        Once we have iterated through every bit of our bitstring, the $j$-invariant of the final elliptic curve $E$ will be the hash value of $b$. 
        
        The rest of this paper will be spent developing tools to study the probability distribution of hash values in the CGL hash. We will also use this information to assess its collision resistance.
        
\section{Isogeny Graphs}\label{Sec:graphs}
    At each step in the CGL hash function, a decision is made about which root to use to keep moving forward. In this way, choosing a bitstring is like choosing a path --- at each bit we decide whether to go left or right. This path can often be quite convoluted, and can cycle back to nodes we have already seen before. Therefore, it would be useful to look at the collection of paths as a whole. With this in mind, we define the concept of an isogeny graph.
    
    \begin{Def}
        The \textbf{complete $l$-isogeny graph} for a field $K$ is a directed pseudograph where the vertices form the set of isomorphism classes of elliptic curves defined over $\bar{K}$, and there is an edge between curves $E$ and $E'$ for every degree $l$ isogeny $\phi: E \to E'$ defined over $\bar{K}$.
    \end{Def}
    
    The reason this is a pseudograph and not a graph is that it is possible for there to be more than one edge/isogeny going from a node $E$ to another node $E'$. It is also possible for isogenies to go from a curve to another curve with the same $j$-invariant, resulting in the graph having self-loops.
    
    \begin{Prop}\cite[Sec.~II.2.11]{Silverman}
        Let $E: y^2 = x^3 + ax + b$ be an elliptic curve defined over a finite field $K$ with $\char{K} = p$. Let $q = p^d$ be some power of $p$. Then, the $q$-Frobenius map $\phi_q: E \to E^{(q)}$, where $E^{(q)}: y^2 = x^3 + a^qx + b^q$, given by $(x,y) \mapsto (x^q, y^q)$ is an inseparable isogeny of degree $q$. 
    \end{Prop}
    
    \begin{DefProp}\cite[Sec.~V.3.1]{Silverman}
        Let $E: y^2 = f(x)$ be an elliptic curve defined over a finite field $K$ with $\char{K} = p$. The following are equivalent:
            \begin{enumerate}
                \item There are no non-trivial $p$-torsion points on $E$ over any algebraic extension of $K$.
                \item The multiplication map $[p]:E \to E$ is not separable.
                \item The coefficient of $x^{p-1}$ in $f(x)^{\frac{p-1}{2}}$ is $0$.
                \item The dual $\hat{\phi}_p$ of the $p$-Frobenius map is inseparable.
            \end{enumerate}
        
        If $E$ satisfies these conditions, then it is said to be \textbf{supersingular}. Otherwise, the curve is called \textbf{ordinary}.
    \end{DefProp}
    
    \begin{Prop}
        Let $\phi: E \to E'$ be an isogeny of degree $l$ defined over a finite field $K$ with $\char{K} = p$ such that $\gcd{p, l} = 1$. Then, $E$ and $E'$ are either both supersingular, or both ordinary. 
    \end{Prop}
    
    \begin{proof}
        Suppose, for the sake of contradiction, that $E$ is supersingular and $E'$ is ordinary. This means that there exists some non-trivial point $Q \in E'$ such that $[p](Q) = \mathcal{O}_{E'}$. In other words, $Q$ is a $p$-torsion point of $E'$. Now, consider the dual map $\hat{\phi}$. 
            $$\phi \circ \hat{\phi}(Q) = [l](Q)$$
        Because $[p](Q) = \mathcal{O}_{E'}$, we see that the order of $Q$ must divide $p$, and is therefore either $1$ or $p$ because $p$ is prime. But we chose $Q$ to not be the identity, and so in fact it must have order $p$. We can reduce $l$ modulo $p$ and write $l \equiv d \mod p$, where $d \in \lrc{0, 1, 2, \dots, p-1}$. Since $\gcd{l,p} = 1$, we see that $d \neq 0$. This means that we can write $l = kp + d$ for some integer $k$. Therefore,
            $$[l](Q) = [k][p](Q) \oplus [d](Q) = [d](Q)$$
        But $[d](Q) \neq \mathcal{O}_{E'}$ because $d$ is smaller than the order of $Q$. We conclude that $\phi \circ \hat{\phi}(Q) \neq \mathcal{O}_{E'}$.
        
        On the other hand $[p](Q) = \mathcal{O}_{E'}$ implies $\hat{\phi}([p](Q)) = \mathcal{O}_E$ because $\hat{\phi}$ is a homomorphism. Pulling out the multiplication by $p$, we get $[p](\hat{\phi}(Q)) = \mathcal{O}_E$. We conclude that $\hat{\phi}(Q)$ is a $p$-torsion point of $E$. But $E$ is supersingular, and so has no non-trivial $p$-torsion points. Therefore, $\hat{\phi}(Q) = \mathcal{O}_{E}$, further implying that $\phi \circ \hat{\phi}(Q) = \mathcal{O}_{E'}$. But just one paragraph ago, we saw that $\phi \circ \hat{\phi}(Q)\neq \mathcal{O}_{E'}$. This is a contradiction.
        
        To complete the proof, we must also rule out the case where $E$ is ordinary and $E'$ is supersingular. But this follows by reversing the roles of $E$ and $E'$ and also those of $\phi$ and $\hat{\phi}$ in the above argument.
    \end{proof}
    
    \begin{Cor}
        Taking $l = 1$ in the above proposition, we see that supersingularity is preserved by isomorphisms.
    \end{Cor}

    \begin{Def}
        We say that a $j$-invariant is supersingular if there exists a supersingular curve with that $j$-invariant. 
    \end{Def}    
    
    The above proposition tells us that when $\gcd{l,p} = 1$, the portion of the graph with supersingular curves never touches that with ordinary curves, and so we might as well consider the cases separately. Further, the portion of the graph that consists of ordinary curves has very rigid, predictable structure (these graphs are often called "volcano" graphs), which makes for poor hash functions. For more details, see \cite[Sec.~25.4]{Galbraith}. Therefore, the rest of this paper will be concerned with supersingular isogeny graphs.
    
    \begin{Def}
        The \textbf{supersingular $l$-isogeny graph} $G_l(K)$ of a finite field $K$ with $\char{K} = p$ is the subgraph of the complete $l$-isogeny graph containing only supersingular curves over $\bar{K}$.
    \end{Def}
    
    In order to generate supersingular isogeny graphs, we need a few extra facts to help computation go smoothly.
    
    \begin{Prop}\label{Prop:Fp2}
        Every supersingular curve over a finite field $K$ with $\char{K} = p$ is isomorphic to a curve that is defined over $\F_{p^2}$. Further, if $E: y^2 = f(x) = x^3 + ax + b$ is supersingular with $a,b \in \F_{p^2}$ and $j(E) \neq 0, 1728$, then the roots of $f(x)$ are also in $\F_{p^2}$.
    \end{Prop}
    
    Before proving this, we need a couple of lemmas.
    
    \begin{Lemma}\cite[Sec.~II.2.12]{Silverman}.
        Let $\psi: E \to E'$ be an isogeny defined over a finite field $K$ with $\char{K} = p$. Then, there exists $q$, a power of $p$, and a separable isogeny $\lambda: E^{(q)} \to E'$ such that $\psi = \phi_q \circ \lambda$.
    \end{Lemma}
    
    \begin{Lemma}\cite[Sec.~V.5.1]{Lang}
        Let $x \in \bar{\F}_p$. Then, $x \in \F_{p^d}$ if and only if $x^{p^d} = x$. 
    \end{Lemma}
    Now, we can return to the proof of \cref{Prop:Fp2}.
    
    \pagebreak
    \begin{proof}
        Let $E: y^2 = x^3 + ax + b$ be supersingular over a finite field $K$ with $\char{K} = p$. We can look at the $p$-Frobenius map $\phi_p$ and its dual $\hat{\phi}_p$. Since $E$ is supersingular, we know that $\hat{\phi}_p$ is inseparable of degree $p$. Therefore, by the lemma, we can factor it as follows:
            $$\begin{tikzcd}
                E \arrow[r, "\phi_p"] \arrow[rrd, "\phi_{p^2}"'] & E^{(p)} \arrow[rd, "\phi_p"] \arrow[rr, "\hat{\phi}_p"] &                                & E \\
                & & E^{(p^2)} \arrow[ru, "\lambda"'] &  
            \end{tikzcd}$$  
        Here, we know that the map from $E^{(p)}$ to $E^{(p^2)}$ must be the $p$-Frobenius map because its degree must divide that of $\hat{\phi}_p$, which is $p$. This further implies that $\deg\lrp{\lambda} = 1$, and so $\lambda \circ \hat{\lambda} = [1] = id$. Therefore, $\lambda$ is invertible with inverse $\hat{\lambda}$, making $\lambda$ an isomorphism.
        
        It follows that $j(E^{(p^2)}) = j(E)$. However,
            \begin{align*}
                j(E^{(p^2)}) &= 1728 \cdot \frac{4a^{3p^2}}{4a^{3p^2} + 27b^{2p^2}} \\
                &\equiv \lrp{1728 \cdot \frac{4a^3}{4a^3 + 27b^2}}^{p^2} \mod p \\
                &= j(E)^{p^2}
            \end{align*}
        because elements of $\F_p$ (in this case $1728, 4, 27$) are fixed when raised to a power of $p$, and because $(a + b)^p \equiv a^p + b^p \mod p$. We conclude that $j(E) = j(E)^{p^2}$, implying that $j(E) \in \F_{p^2}$. We must show that this implies the existence of a curve isomorphic to $E$ but defined over $\F_{p^2}$. Given $j(E) = j \in \F_{p^2}$, such a curve can be constructed as
            $$y^2 = x^3 + \frac{3j}{1728 - j} \cdot x + \frac{2j}{1728 - j}$$
        \cite[Sec.~2.1]{Adj} To see that this curve does indeed have $j$-invariant $j$, we simply plug the coefficients into the formula, after which simple algebraic manipulation yields the desired result. We note that this formula does not work when $j = 1728$ because of a division by $0$, or when $j = 0$, in which case the curve given by the formula has discriminant $\Delta = 0$. In such cases, we simply use curves of the form $y^2 = x^3 + ax$ and $y^2 = x^3 + b$ respectively, with $a, b \in \F_{p^2}$ non-zero.
        
        Now, suppose that $E: y^2 = x^3 + ax + b$ is supersingular with $a,b \in \F_{p^2}$ and $j(E) \neq 0, 1728$. Let $x_0$ be a root of $x^3 + ax + b$, so that $(x_0, 0)$ is a point of order $2$ on $E$. Then,
            $$\begin{tikzcd}
                {(x_0, 0)} \arrow[r, "\phi_p", maps to] \arrow[rrd, "\phi_{p^2}"', maps to] & {(x_0^p, 0)} \arrow[rd, "\phi_p", maps to] \arrow[rr, "\hat{\phi}_p", maps to] &                                                  & {[p](x_0, 0)} \\
                 & & {(x_0^{p^2}, 0)} \arrow[ru, "\lambda"', maps to] &              
            \end{tikzcd}$$
        
        \pagebreak
        We consider the curve $E^{(p^2)}: y^2 = x^3 + a^{p^2}x + b^{p^2}$. Since $a,b \in \F_{p^2}$, we see that $a^{p^2} = a$ and $b^{p^2} = b$, implying that $E^{(p^2)} = E$. Therefore, $\lambda$ is actually an automorphism of $E$. But since $j(E) \neq 0, 1728$, we have that $\aut{E} = \lrc{id, [-1]}$, where $[-1](x,y) = (x,-y)$ for any point $(x,y) \in E$ \cite[Sec.~III.10.1]{Silverman}. Since neither of these automorphisms affect points of the form $(x,0)$, we conclude that $\lambda(x_0^{p^2}, 0) = (x_0, 0)$, implying that $x_0^{p^2} = x_0$. Thus, $x_0 \in \F_{p^2}$, completing the proof.
    \end{proof}
    
    Note, the proof does not work for nodes with $j$-invariant $0$ or $1728$ because such curves have larger automorphism groups containing elements that might not all fix points of the form $(x, 0)$. In fact, there are many curves with these $j$-invariants, defined over $\F_{p^2}$ such that $f(x)$ does not have all three roots in $\F_{p^2}$.
    
    \begin{Rem}\label{Rem:Fp2}
        The first part of the above proposition allows us to look at isogeny graphs over $\F_p^2$ instead of over $\bar{\F_P}$, which greatly reduces the amount of computation required. This is another big reason why supersingular graphs make for better hash functions --- they are much faster to compute. This also tells us that there are only finitely many vertices in the graph, because there are only finitely many curves defined over $\F_{p^2}$, and only a subset of those are supersingular.
    \end{Rem}
    
    \begin{Rem}\label{Rem:veluFp2}
        The second part of the proposition shows us another big advantage of using supersingular curves. If $E: y^2 = f(x) = x^3 + ax + b$ is a supersingular elliptic curve defined over $\F_{p^2}$, then we might ask whether the new curves produced as co-domains of isogenies obtained from V\'{e}lu's formulae are also themselves defined over $\F_{p^2}$, as opposed to just being isomorphic to curves defined over $\F_{p^2}$. We note that because V\'{e}lu's formulae only use field operations on $a,b$ and a root $x_0$ of $f(x)$, the codomain curve will be defined over $\F_{p^2}$ if $x_0$ is in $\F_{p^2}$. This is exactly what the proposition gives us, at least for curves of $j$-invariant $\neq 0, 1728$. 
    \end{Rem}
    
    \begin{Th}\label{Thm:connected}\cite[Corollary~ 78]{Kohel}
        Given $K$, a finite field with $\char{K} = p$, the graph $G_l(K)$, with $l \neq p$ prime, is connected.
    \end{Th}
    
    With the above facts at our disposal, we can now create an algorithm to generate the isogeny graphs $G_2(K)$ for any finite field $K$. The algorithm, which can be found implemented in SageMath at \href{https://github.com/dhruvbhatia00/CGL-Hash.git}{https://github.com/dhruvbhatia00/CGL-Hash.git}, works as follows. Given a prime number $p$, we start by finding a supersingular elliptic curve $E$ defined over $\F_{p^2}$. The $j$-invariant of $E$ will be the first node of our graph. We also create a queue $Q$ containing $E$. We repeat the following, in order, for each element $N$ of the queue, until it is empty:
    
    \begin{enumerate}
        \item Write $N: y^2 = f(x)$ and compute the three roots $x_1, x_2, x_3$ of $f(x)$.
        \item Use V\'{e}lu's formulae to compute three isogenies $\phi_1, \phi_2, \phi_3$, each corresponding to the kernels $\lrc{\mathcal{O}_N, (x_1, 0)}$, $\lrc{\mathcal{O}_N, (x_2, 0)}$, $\lrc{\mathcal{O}_N, (x_3, 0)}$ respectively. Let the corresponding codomains be $E_1, E_2, E_3$.
        \item We compute the $j$-invariant for each of $E_1, E_2, E_3$, and for every $j$-invariant we encounter for the first time, we add a new node to the graph. We also add in arrows representing each of the three isogenies. Among $E_1, E_2, E_3$, those with new $j$-invariants are added to the end of $Q$.
    \end{enumerate}
    
    Since there are only finitely many supersingular curves over a finite field, this algorithm must terminate. At each step, \cref{Prop:Fp2} ensures that the new curves stay defined over $\F_{p^2}$, as described in \cref{Rem:Fp2} and \cref{Rem:veluFp2}. The only time this might not be the case is when $j=0$ or $j=1728$. Fortunately, there is an easy fix. 
    
    Let $E: y^2 = x^3 + ax + b$ have $j$-invariant $0$. So,
        $$j(E) = 1728 \cdot \frac{4a^3}{4a^3 + 27b^2} = 0$$
    We conclude that $a = 0$, and so $E: y^2 = x^3 + b$. We can see further that irrespective of what $b$ is, when $a = 0$, $j(E) = 0$. This means, in particular, that the $j$-invariant $0$ can be represented over $\F_{p}$ by the curve $E: y^2 = x^3 - 1 = (x-1)(x^2 + x + 1)$. Here, since $x^2 + x + 1 \in \F_p[x]$, its roots will necessarily exist over $\F_{p^2}$.
    
    Similarly, all curves with $j$-invariant $1728$ can be represented by a curve of the form $E: y^2 = f(x) = x^3 + ax = x(x^2 + a)$. So, as long as we choose $a \in \F_p$ (for example, our algorithm chooses $a = 1$), all roots of $f(x)$ will be in $\F_{p^2}$.
    
    This means that every time the current node has $j$-invariant $0$ or $1728$, we can simply use representative curves as above, and still be sure that we never leave $\F_{p^2}$. Finally, \cref{Thm:connected} ensures that this algorithm reaches all supersingular $j$-invariants over $\F_{p^2}$.

    In our study of supersingular isogeny graphs, it is useful to know how many vertices the graph has. In other words, we would like to know how many curves are supersingular over a finite field with characteristic $p$.
    \begin{Prop}\cite[Sec.~V.4.1]{Silverman}
        The number of supersingular curves up to isomorphism over $\bar{\F}_p$ is
            \[ 
                \floor{\frac{p}{12}} + \begin{cases}
                    0 & \text{if $p \equiv 1 \pmod{12}$} \\
                    1 & \text{if $p \equiv 5 \pmod{12}$} \\
                    1 & \text{if $p \equiv 7 \pmod{12}$} \\ 
                    2 & \text{if $p \equiv 11 \pmod{12}$}
                 \end{cases}
            \]
    \end{Prop}
    
    We now take a look at some examples of supersingular isogeny graphs over different fields.
    
    \begin{Ex}
        Let $K = \F_{61}$ be the field with $61$ elements. Since $p = 61 \equiv 1 \mod 12$, we should expect to see $\floor{\slfrac{61}{12}} = 5$ nodes.
        \begin{figure}[H]
            \centering
            \includegraphics[scale=.8]{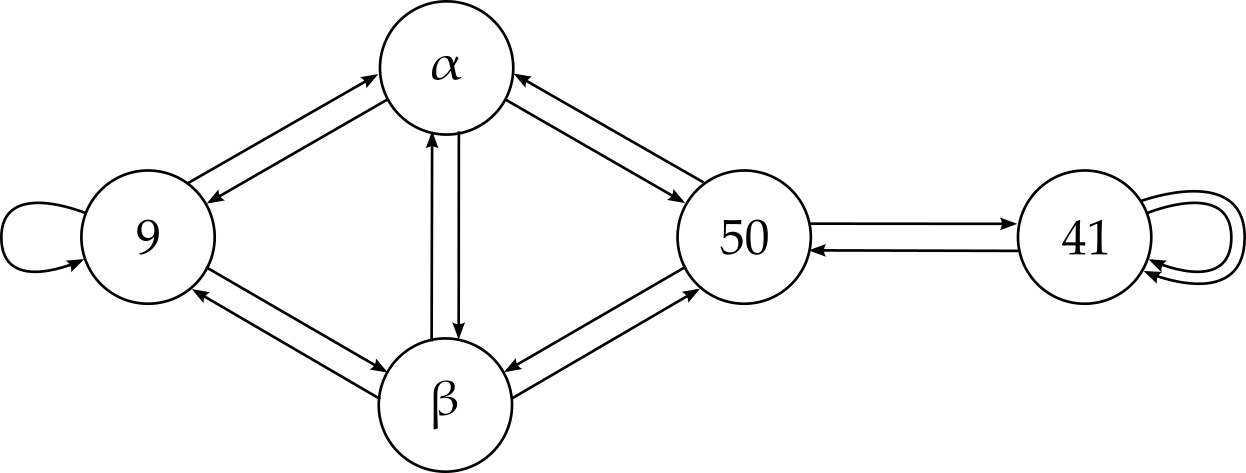} 
            \caption{$G_2(\F_{61})$}
            \label{fig:p=61}
        \end{figure}
        All nodes above (\cref{fig:p=61}) are labelled with their $j$-invariants. Since all the curves are supersingular, we know that all the $j$-invariants are elements of $\F_{61^2}$. Here, $\alpha = 20z + 32$ and $\beta = 41z + 52$, where $z \in \F_{61^2}$ is a root of $x^2 + 60x + 2$ over $\F_{61}$. We note the graph is completely $3$-regular (each node has three edges entering and leaving it), and every arrow has a dual, as expected.
    \end{Ex}
    
    \begin{Ex}
        Let $K = \F_{41}$. This time, $p = 41 \equiv 5 \mod 12$, and so we expect there to be $\floor{\slfrac{41}{12}} + 1 = 4$ nodes.
        \begin{figure}[H]
            \centering
            \includegraphics[scale=.8]{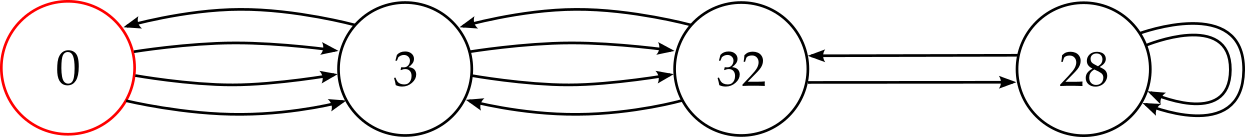} 
            \caption{$G_2(\F_{41})$}
            \label{fig:p=41}
        \end{figure}
        In \cref{fig:p=41}, all nodes exhibit $3$-regular behaviour except $j=0$ (highlighted in red) and its neighbour $j=3$. Somehow, there seem to be three arrows out of $0$ (all going to $3$), but only one arrow into $0$ from $3$.
    \end{Ex}
        
    \begin{Ex}   
        Let $K = \F_{43}$. By the counting formula, we should expect four nodes because $p = 43 \equiv 7 \mod 12$.
        \begin{figure}[H]
            \centering
            \includegraphics[scale=.8]{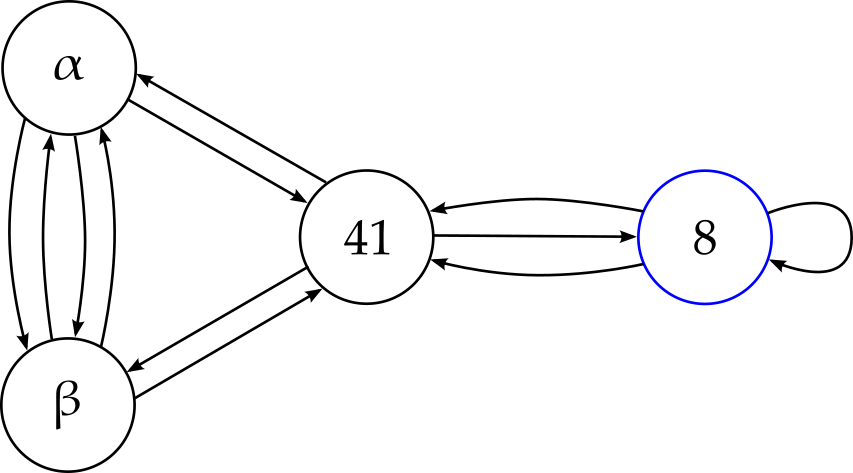} 
            \caption{$G_2(\F_{43})$}
            \label{fig:p=43}
        \end{figure}
        Here, in \cref{fig:p=43}, $\alpha = 39z + 14$ and $\beta = 4z + 10$, where $z \in \F_{43^2}$ is a root of $x^2 + 42x + 3$ over $\F_{43}$.
        This time, the problem node seems to be $j = 8$, which we point out is congruent to $1728 \mod 43$. This node has three arrows going out, but only two going in.
    \end{Ex}
    
    \begin{Ex}
        Let $K = \F_{47}$. The counting formula implies that we should see five nodes because $p = 47 \equiv 11 \mod 12$.
        \begin{figure}[H]
            \centering
            \includegraphics[scale=.8]{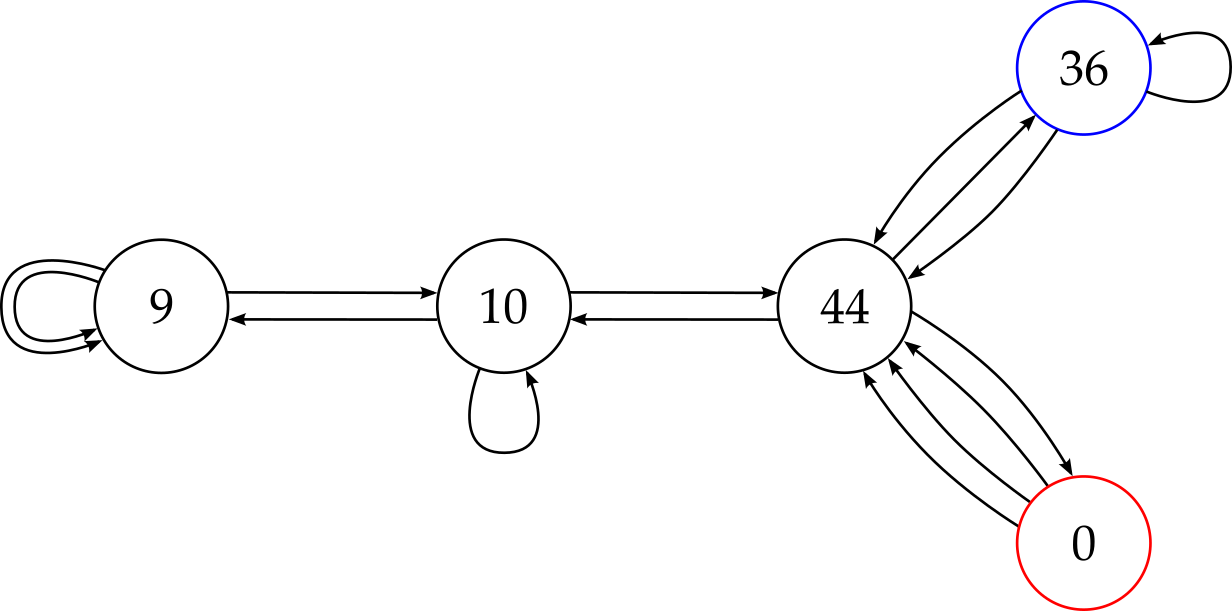} 
            \caption{$G_2(\F_{47})$}
            \label{fig:p=47}
        \end{figure}
        \Cref{fig:p=47} has two problem nodes: $j = 0$ and $j = 36 \equiv 1728 \mod 47$.
    \end{Ex}
    
    \begin{Rem}
        In the above examples, we saw the problem nodes have more arrows to their neighbours than there are arrows going back. For example, $j=0$ always seems to have three arrows pointing at its neighbour, but only one arrow back. This should seem impossible because every isogeny comes with a unique dual isogeny in the other direction. We remind the reader, however, that duals are only unique up to post-composition by an automorphism. So, we conclude that problem nodes like $j = 0$ must have extra automorphisms making all three arrows together be duals of the single arrow in the other direction. This is discussed in more detail in \cref{Sec:distributions}.
    \end{Rem}
    
    In the above examples, $j=0$ and $j=1728$ seemed to have strange behaviour. To better study this, it would be useful to know when these $j$-invariants are supersingular.
    
    \begin{Prop}
        Let $K$ be a finite field with $\char{K} = p$. The $j$-invariant $j = 0$ is supersingular if and only if $p \equiv 2 \mod 3$.
    \end{Prop}
    
    \begin{proof}
        Let $E: y^2 = f(x) = x^3 + b$ be an elliptic curve with $j(E) = 0$. Checking whether this curve is supersingular amounts to checking whether the coefficient of $x^{p-1}$ in $f(x)^{\frac{p-1}{2}} = (x^3 + b)^{\frac{p-1}{2}}$ is $0$ in $K$.
        
        \pagebreak
        We can use the binomial theorem to find out what the $x^{p-1}$ term looks like. Each term in the expansion of $(x^3 + b)^{\frac{p-1}{2}}$ is of the form
            $$\binom{\frac{p-1}{2}}{k} (x^3)^k \cdot b^{\frac{p-1}{2} - k}$$
        where $0 \leq k \leq \frac{p-1}{2}$ is an integer. So, to get the $x^{p-1}$ term, we need $k = \frac{p-1}{3}$. But $\frac{p-1}{3}$ is an integer if and only if $p \equiv 1 \mod 3$. We conclude that when $p \equiv 2 \mod 3$, there is no $x^{p-1}$ term in $f(x)^{\frac{p-1}{2}}$, and so $E$ is supersingular.
        
        On the other hand, when $p \equiv 1 \mod 3$, the term in question is
            $$\binom{\frac{p-1}{2}}{\frac{p-1}{3}} \cdot \lrp{x^3}^{\frac{p-1}{3}} \cdot b^{\frac{p-1}{2} - \frac{p-1}{3}}$$
        Here, because $b \in K$ is non-zero, we see we only care about the binomial coefficient
            $$\binom{\frac{p-1}{2}}{\frac{p-1}{3}} = \frac{\lrp{\frac{p-1}{2}}!}{\lrp{\frac{p-1}{3}}! \cdot \lrp{\frac{p-1}{6}}!}$$
        Since everything in both the numerator and denominator is $>0$ and $< p$, we see that the coefficient is not $0$, proving that the curve is not supersingular, as needed.
    \end{proof}
    
    \begin{Prop}
        Let $K$ be a finite field with $\char{K} = p$. The $j$-invariant $j = 1728$ is supersingular if and only if $p \equiv 3 \mod 4$.
    \end{Prop}
    
    \begin{proof}
        Let $E: y^2 = f(x) = x^3 + ax$ be an elliptic curve with $j(E) = 1728$. Checking whether this curve is supersingular amounts to checking whether the coefficient of $x^{p-1}$ in $f(x)^{\frac{p-1}{2}} = (x^3 + ax)^{\frac{p-1}{2}}$ is $0$ in $K$.
        
        As in the previous proposition, we can use the binomial theorem to find out what the $x^{p-1}$ term looks like. This time, however, things are slightly more complicated. In order to get an $x^{p-1}$ term, there must be some integer $0 \leq k \leq \frac{p-1}{2}$ such that
            $$x^{p-1} = (x^3)^k \cdot x^{\frac{p-1}{2} - k}$$
        We can rewrite this as:
            \begin{align*}
                p-1 = 3k + \frac{p-1}{2} - k \\
                \frac{p-1}{2} = 2k
            \end{align*}
        The only solution for $k$ is $k = \frac{p-1}{4}$, which is only an integer when $p \equiv 1 \mod 4$. When $p \equiv 3 \mod 4$, we see that there is no $x^{p-1}$ term, making the curve supersingular.
        
        But when $p \equiv 1 \mod 4$, we see that the $x^{p-1}$ term is
            $$\binom{\frac{p-1}{2}}{\frac{p-1}{4}} \cdot \lrp{x^3}^{\frac{p-1}{4}} \cdot (ax)^{\frac{p-1}{2} - \frac{p-1}{4}}$$
        Once again, because $a \in K$ is non-zero, we see we only care about the binomial coeffient
            $$\binom{\frac{p-1}{2}}{\frac{p-1}{4}} = \frac{\lrp{\frac{p-1}{2}}!}{\lrp{\frac{p-1}{4}}! \cdot \lrp{\frac{p-1}{4}}!}$$
        Since everything in both the numerator and denominator is $>0$ and $< p$, we see that the coefficient is not $0$, proving that the curve is not supersingular, as needed.
    \end{proof}
    
    \begin{Rem}
        The forward directions of the above two propositions were first proved in \cite{Munuera}.
    \end{Rem}
    
    \begin{Rem}
        Combining the above two propositions, we get four cases working mod $12$. When $p \equiv 1 \mod 3$ and $p \equiv 1 \mod 4$, implying that $p \equiv 1 \mod 12$, we see that neither $j=0$ or $j=1728$ is supersingular. Similarly, if $p \equiv 5 \mod 12$, then $j=0$ is supersingular and $j=1728$ is not. When $p \equiv 7 \mod 12$, $j=1728$ is supersingular and $j=0$ is not. Finally, when $p \equiv 11 \mod 12$, both $j=0$ and $j=1728$ are supersingular.
    \end{Rem}

\section{Stochastic Matrices}\label{Sec:stochastic}
    One of the things we'd like to know about the CGL hash is how likely it is for two randomly chosen bitstrings to collide at the same hash value. One way of computing this would be to first compute the probability of a randomly chosen bitstring attaining a specified hash value. In other words, we would be computing a probability distribution for all the hash values. In this section, we describe a method of using stochastic matrices to represent isogeny graphs from which we can compute these probability distributions. 
    
    \begin{Def}
        A \textbf{left stochastic matrix} is a square matrix $M$ with non-negative real entries such that the sum of values in each column is $1$.
    \end{Def}

    Let $G_2(K)$ be the isogeny graph of supersingular elliptic curves over a finite field $K$ of characteristic $p>3$. We would like to construct an $n \times n$ matrix $M$, where $n$ is the number of vertices in $G_2(K)$ such that the entry in the $i^\text{th}$ column and $j^\text{th}$ row corresponds to the probability of moving from the $i^\text{th}$ node to the $j^\text{th}$ node in the graph. Unfortunately, this is not so simple because the probability of moving from the $i^\text{th}$ node to the $j^\text{th}$ node depends on where we arrived at the $i^\text{th}$ node.
    
    \pagebreak
    
    Recall that at each step in the hash function, we compute the three roots, get rid of the root corresponding to the dual of isogeny we just used, and then choose one of the remaining roots based on what the current bit is. Given a random bitstring, this bit has a $0.5$ chance of being a $0$ and a $0.5$ chance of being a $1$, implying that both remaining roots are equally likely to get chosen, while the first root (the one we got rid of), has $0$ chance of being chosen because we disallow backtracking. However, we cannot know which root we just got rid of without taking into account where we came to the current node from. So, we look at each current and previous node pair separately.
    
    We make a matrix $M$ with a row and column for every valid ordered pair $(E, E')$ of nodes in the graph, where a pair $(E, E')$ is called valid if there is an arrow $E' \to E$ in the graph. The first element of each valid pair represents the current node, and the second element represents the previous node. In $M$, we fill the spot at column $(E_0, E_{0}')$ and row $(E_1, E_{1}')$ with the probability of moving from $E_0$ (having just come from $E_{0}'$) to $(E_1)$ (having just come from $E_{1}'$). Clearly, this will only be non-zero if $E_0 = E_1'$, so that after moving from $E_0$ to $E_1$, the current node is $E_1$, and the previous node is $E_0$.
    
    We will go through an example when $p=23$ to illustrate this. This graph has three nodes with $j$-invariants $0,19,$ and $1728$.
    \begin{figure}[H]
        \centering
        \includegraphics[scale=.8]{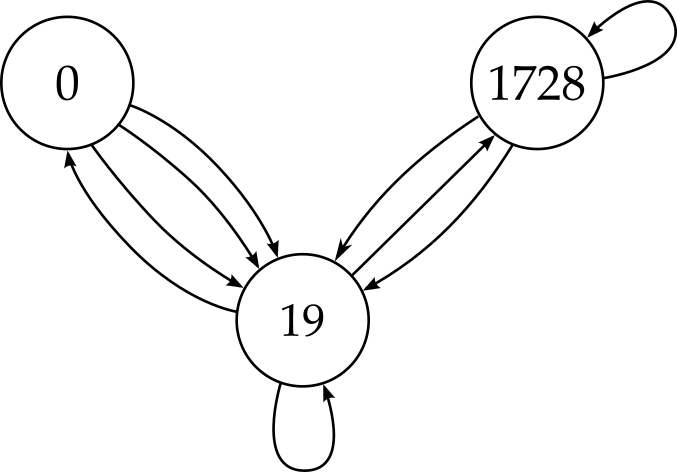} 
        \caption{$G_2(\F_{23})$}
        \label{fig:p=23}
    \end{figure}

    We label each column with an ordered pair $(a,b)$, representing the current and the previous node respectively, likewise with the rows. Starting at column $(19,0)$, meaning we are currently at node with $j$-invariant $19$, having just come from $0$, we write in the respective row the probabilities of going to that node next. As we can see in \cref{fig:p=23}, 
    node $19$ has three outward edges, one going to each of the three nodes. However, since we were just at $0$, we cannot go back because the isogeny from $0$ to $19$ is dual to the isogeny from $19$ to $0$. So, we either go to $1728$ or self-loop back to $19$ with equal probability. We will denote this with a $.5$ in both rows $(19,19)$ and $(1728,19)$. We continue filling in the columns in this manner. Since each column has probabilities which necessarily add up to $1$, it is a left stochastic matrix.

    \renewcommand{\kbldelim}{(}
    \renewcommand{\kbrdelim}{)}
    $$\kbordermatrix{
            &{(19,0)} & {(19,19)} & {(19,1728)} & {(0,19)} & {(1728,19)} & {(1728,1728)}\\
            (19,0) &   0 & 0 & 0 & 1 & 0 & 0 \\
            (19,19) & 0.5 & 0 & 0.5 & 0 &0 & 0 \\
            (19,1728) &  0 & 0 & 0 & 0 & 0.5 & 1 \\
            (0,19) & 0 & 0.5 & 0.5 & 0 & 0 & 0 \\
            (1728,19) & 0.5 & 0.5 & 0 & 0 & 0 & 0 \\
            (1728,1728) &0 & 0 & 0 & 0 & 0.5 & 0 \\
            }$$
    
    Suppose that in our hash function, we decide to start at the node $19$ and a root $x_1$ that corresponds to an isogeny to the node $0$. In other words, we are starting at the pair $(19, 0)$. We can represent this state with a vector $\vec{v}$ with $1$ in the entry corresponding to $(19, 0)$, and $0$s everywhere else. We see then that the probabilities of being at a pair after one bit are represented by the vector $M \cdot \vec{v}$. After two bits, the probabilities are represented by $M \cdot (M \cdot \vec{v}) = M^2 \cdot \vec{v}$. In general, after $n$ bits, the probabilities of being at a certain node pair are represented by $M^n \cdot \vec{v}$. In this example, we notice that as we increase $n$, the vector $M^n \cdot \vec{v}$ seems to be approaching
        $$\kbordermatrix{
         & \\
        (19,0) & \slfrac{2}{11}  \\
        (19, 19) & \slfrac{2}{11}  \\
        (19, 1728) & \slfrac{2}{11}  \\
        (0, 19) & \slfrac{2}{11}  \\
        (1728, 19) & \slfrac{2}{11}  \\
        (1728, 1728) & \slfrac{1}{11}
        }$$
    So, for a sufficiently long bitstring, the above values give a good approximation for the probability of being at a certain (current, previous) pair. If we want to find the probability of being at a certain node, we simply add up the entries of all pairs with that current node. In this example, those probabilities are:
        \begin{align*}
            P(19) &= 3 \cdot \frac{2}{11} = \frac{6}{11}\\
            P(0) &= \frac{2}{11}\\
            P(1728) &= \frac{2}{11} + \frac{1}{11} = \frac{3}{11}
        \end{align*}
    
    We can see that if $M^n \cdot \vec{v}$ converges to some vector $\vec{x}$, then it must be the case that $M \cdot \vec{x} = \vec{x}$, implying that $\vec{x}$ is an eigenvector of $M$ with eigenvalue $1$. Indeed, the vector with $\frac{2}{11}$ and $\frac{1}{11}$ in the appropriate positions is such an eigenvector for this example.
    
    \begin{Th}\cite[Chp 4.9 Thm.~18]{Lay}
        Every left stochastic matrix $M$ has an eigenvector with eigenvalue $1$ such that if $\vec{v}$ is a vector representing a probability distribution (its entries are non-negative reals that add to $1$), then $\lim_{n \to \infty} M^n \cdot \vec{v}$ is this eigenvector. 
    \end{Th}
    
    In order to find the probability distribution of hash values for a sufficiently long bitstring, all we need to do is compute the eigenvector with eigenvalue $1$, scale it appropriately so that its values sum to $1$, and then sum entries by current node.

\section{Expected Probability Distribution}\label{Sec:distributions}
    In this section, we construct the expected probability distributions based on our data gathered from the stochastic matrices. Then, we prove the probabilities of random hash values approach these distributions. 
    

    \begin{Th}\label{Thm:main} 
        Let $E$ be a supersingular elliptic curve with $j$-invariant $j$ over a field $F_{p^2}$ and $p>3$ a prime. Then the probability of a sufficiently long bitstring $b$ having a hash value equal to $j$ approaches:
            \[
                P(j) =
                \begin{cases}
                    \dfrac{6}{\frac{p-1}{2}} & \text{if $j \not \equiv 1728, 0\pmod{p}$} \\\\
                    \dfrac{3}{\frac{p-1}{2}} & \text{if $j \equiv 1728\pmod{p}$} \\\\
                    \dfrac{2}{\frac{p-1}{2}} & \text{if $j \equiv 0\pmod{p}$}
                \end{cases}
            \]
    \end{Th}
    
    \begin{Rem}
        One might notice that the theorem makes no reference to what $p$ is modulo $12$. Since congruence modulo $12$ is a big part of what determines how many curves are in the graph, it might seem surprising that all the probabilities involved have the same denominator, irrespective of what $p$ is mod $12$. To dispel some of these fears, we include the following computations.
        
        When $p \equiv 1 \mod 12$, both $j=0$ and $j=1728$ are not supersingular, and so there are $\frac{p-1}{12}$ nodes, each with probability $\slfrac{6}{\frac{p-1}{2}}$. Adding these together, we get
            $$\frac{p-1}{12} \cdot \dfrac{6}{\frac{p-1}{2}} = 1$$
        When $p \equiv 5 \mod 12$, we see that $j=0$ is supersingular, in addition to $\frac{p-5}{12}$ other supersingular nodes with $j \neq 0, 1728$. Adding this together, we get
            $$\frac{p-5}{12} \cdot \dfrac{6}{\frac{p-1}{2}} + \dfrac{2}{\frac{p-1}{2}} = \frac{p-5}{p-1} + \frac{4}{p-1} = 1$$
        When $p \equiv 7 \mod 12$, we have that $j=1728$ is supersingular, in addition to $\frac{p-7}{12}$ other supersingular curves. Adding,
            $$\frac{p-7}{12} \cdot \dfrac{6}{\frac{p-1}{2}} + \dfrac{3}{\frac{p-1}{2}} = \frac{p-7}{p-1} + \frac{6}{p-1} = 1$$
        Finally, when $p \equiv 11 \mod 12$, we have that both $j=0$ and $j=1728$ are supersingular, along with $\frac{p-11}{12}$ other curves. Adding, we get
            $$\frac{p-11}{12} \cdot \dfrac{6}{\frac{p-1}{2}} + \dfrac{2}{\frac{p-1}{2}} + \dfrac{3}{\frac{p-1}{2}} = \frac{p-11}{p-1} + \frac{4}{p-1} + \frac{6}{p-1} = 1$$
        These computations verify that the values described in the theorem actually do give us probability distributions.
    \end{Rem}
    
    To prove the theorem, we start by proving some lemmas.
    
    \begin{Lemma}\label{lemma:j=0}
        Let $K$ be a finite field with $\char{K} = p > 3$. Suppose that $p \equiv 2 \mod 3$ so that $j = 0$ is supersingular. Then, all three isogenies out of the node $j = 0$ in $G_2(K)$ are equivalent up to pre-composition of an automorphism, even though their kernels are not the same.
    \end{Lemma}
    
    \begin{proof}
        Consider the set $\lrc{\phi_1, \phi_2, \phi_3}$ of separable degree $2$ isogenies with domain $E: y^2 = f(x) = x^3 + b$. Each $\phi_i$ has a kernel $\lrc{\mathcal{O}_E, (x_i, 0)}$, where $x_i$ is a root of $f(x)$. We wish to show that for each pair $i,j$ with $i \neq j$ and $i,j \in \lrc{1,2,3}$, there is an automorphism $\lambda: E \to E$ such that $\phi_j$ and $\lambda \circ \phi_i$ have the same kernel. Equivalently, for each pair $i,j$ with $i \neq j$ and $i,j \in \lrc{1,2,3}$, we wish to show that there exists an automorphism $\lambda: E \to E$ such that $\lambda(x_i,0) = (x_j,0)$.
        
        We observe that any automorphism takes order $2$ points to order $2$ points, and therefore the automorphism group $\aut{E}$ acts on the set $\lrc{(x_1, 0), (x_2, 0), (x_3, 0)}$ of order $2$ elements. Reframing the problem in the language of group actions, we wish to show that this action is transitive. 
        
        As with any group action, there is a group homomorphism $\pi: \aut{E} \to S_3$ such that given $\lambda \in \Aut{E}$ and $i \in \lrc{1,2,3}$, we have that $\lambda(x_i,0) = (x_{\pi(\lambda)(i)},0)$. In \cite[Sec.~III.10.1]{Silverman}, we see that the automorphism group for a curve $E$ over $K$ with $j(E) = 0$ is cyclic with order $6$. Let $\lambda$ be a generator for this group. We note that for every elliptic curve, the map $[-1]$ is an automorphism that fixes order $2$ elements. $[-1]$ has order $2$ in $\aut{E}$, and so $[-1] = \lambda^3$. But $[-1]$ fixes order $2$ elements of $E$, and so $\pi(\lambda^3) = id \in S_3$. This further implies that $\pi(\lambda)^3 = id$. We are left with two possibilities: either $\pi(\lambda) = id$ or $\pi(\lambda)$ is a $3$-cycle $(1 2 3)$ or $(3 2 1)$. In the latter case, if $\pi(\lambda)$ is a $3$-cycle, we see that we can get from any order $2$ element to another by simply applying $\lambda$ or $\lambda^{-1}$, making the action transitive. 
        
        Therefore, we now must rule out the possibility that $\pi(\lambda) = id$. To do this, we look more closely at the automorphisms involved. \cite[Sec.~III.10.1]{Silverman} tells us that automorphisms of a curve $E$ with $j(E)=0$ are of the form
            \begin{align*}
                x &\mapsto u^2x \\
                y &\mapsto u^3y
            \end{align*}
        where $u^6 = 1$. So, without loss of generality, we can assume $\lambda$ is the map that uses $u = a$, where $a$ is a primitive $6^\text{th}$ root of unity (so that $\lambda$ generates $\aut{E}$ the same way $a$ generates the group of $6^\text{th}$ roots of unity). Now, suppose that $\pi(\lambda) = id$. Then,
            $$\lambda(x_i, 0) = (a^2x_i, 0) =(x_i, 0)$$
        for all $i$. But this is impossible unless $x_i = 0$ for all $i$. Since $f(x) = x^3 + b$, where $b$ is non-zero, $0$ cannot be a root of $f$. 
    \end{proof}
    
    \begin{Lemma}\label{lemma:j=1728}
        Let $K$ be a finite field with $\char{K} = p > 3$. Suppose that $p \equiv 3 \mod 4$ so that $j = 1728$ is supersingular. Then, two of the three isogenies out of the node $j = 1728$ in $G_2(K)$ are equivalent up to pre-composition of an automorphism, even though their kernels are not the same. The third isogeny is a self-loop.
    \end{Lemma}
    
    \begin{proof}
        We start by setting things up as in the previous lemma. Given a curve $E: y^2 = f(x) = x^3 + ax$ (so that $j(E) = 1728$), each separable degree $2$ isogeny $\phi_i$ with domain $E$ has kernel $\lrc{\mathcal{O}_E, (x_i, 0)}$, where $x_i$ is a root of $f(x)$. We see that $f(x) = x(x^2 + a)$, and so we can relabel in order to make $x_1 = 0$. As before, $\aut{E}$ acts on the set of order $2$ points of $E$. This time, our goal is to prove that elements of $\aut{E}$ all fix $(x_1, 0) = (0,0)$, while some element swaps $(x_2, 0)$ and $(x_3, 0)$.  
        
        Again, we study the homomorphism $\pi: \aut{E} \to S_3$ such that given $\lambda \in \Aut{E}$ and $i \in \lrc{1,2,3}$, we have that $\lambda(x_i,0) = (x_{\pi(\lambda)(i)},0)$. In \cite[Sec.~III.10.1]{Silverman}, we see that the automorphism group for a curve $E$ over $K$ with $j(E) = 1728$ is cyclic with order $4$. Let $\lambda$ be a generator for this group. As before, $[-1]$ has order $2$ in $\aut{E}$, and so $\lambda^2 = [-1]$. Once again, $\pi(\lambda)^2 = \pi(\lambda^2) = \pi([-1]) = id$. Since $\pi(\lambda)$ is a $2$-torsion point in $S_3$, it must either be $id$ or a $2$-cycle. We will show that in fact $\pi(\lambda)$ must be the $2$-cycle$ (2 3)$.
        
        \cite[Sec.~III.10.1]{Silverman} tells us that automorphisms of a curve $E$ with $j(E)=1728$ are of the form
            \begin{align*}
                x &\mapsto u^2x \\
                y &\mapsto u^3y
            \end{align*}
        where $u^4 = 1$. Without loss of generality, we can assume $\lambda$ is the map that uses $u = a$, where $a$ is a primitive $4^\text{th}$ root of unity. Then,
            $$\lambda(x_i, 0) = (u^2x_i, 0)$$
        Since $x_1 = 0$, we see that $(x_1, 0)$ is fixed by $\lambda$. Since $E$ is an elliptic curve, its discriminant is non-zero, and so $f(x)$ has no repeated roots. This implies that $x_2$ and $x_3$ are non-zero, and so are not fixed by $\lambda$, implying that they are swapped by $\lambda$.
        
        Finally, to see that $\phi_1$ is a self-loop, we can simply plug $x_1 = 0$ into V\'{e}lu's formulae and verify that the new curve produced still has $j$-invariant $1728$. We recall the new curve is given by $\Tilde{E}: y^2 = x^3 + Ax + B$, where
            \begin{align*}
                A &= -15x_1^2 - 4a  = -4a\\
                B &= 14x_1^3 = 0
            \end{align*}
        Since this new curve lacks a constant term, it too has $j$-invariant $1728$, completing the proof.
    \end{proof}
    
    We now return to the proof of our theorem.
    \pagebreak
    
    \begin{proof}
        We recall that we found the probability of landing at a node with $j$-invariant $j$ by first computing the eigenvector associated to eigenvalue $1$ for the matrix $M$ associated with our isogeny graph and then summing up all the entries with current node $j$. So, the goal is to find this eigenvector and show that its entries sum to produce the results described in the theorem. 
        
        Let $P(j_1, j_2)$ describe the probability of arriving at the node with $j$-invariant $j_1$ from the node with $j$-invariant $j_2$. This corresponds to the entry of the eigenvector in the row for $(j_1, j_2)$. We will show that the following values form the eigenvector:
            \[
                P(j_1, j_2) =
                \begin{cases}
                    \dfrac{1}{\frac{p-1}{2}} & \text{if $j_1 = j_2 \equiv 1728 \pmod{p}$} \\\\
                    \dfrac{2}{\frac{p-1}{2}} & \text{if $j_1$ or $j_2 \not\equiv 1728\pmod{p}$ with one dual pair in between} \\\\
                    \dfrac{4}{\frac{p-1}{2}} & \text{if $j_1$ or $j_2 \not\equiv 1728\pmod{p}$ with two dual pairs in between}
                \end{cases}
            \]
        Here, dual pairs refer to an isogeny along with its dual isogeny. When $j=0$ for example, even though there are three arrows from it to its neighbour (as described in \cref{lemma:j=0}), they all have the same dual, and so there is only one dual pair between the nodes. A similar statement can be said about the two isogenies from $j=1728$ to its neighbour. We also note that it is impossible for there to be three dual pairs between any two nodes. If this were the case, these two nodes would be disconnected from the rest of the graph, which is not possible because isogeny graphs for supersingular curves are connected. Still, it might be the case that there are only two nodes with three dual pairs in between. By the supersingular curve counting formula, this can only happen when $p = 11, 17, 19, 25$. When $p \equiv 11, 17$ or $19$, at least one node is of $j=0$ or $j=1728$, which, by \cref{lemma:j=0} and \cref{lemma:j=1728} can never have three dual pairs with a neighbour. Finally, $p = 25$ is not a prime number, and so we conclude that three dual pairs is never possible.
        
        A few quick calculations show that a vector with entries $P(j_1, j_2)$ does indeed give us probabilities as described in the theorem. If a node has $j$-invariant $0$ then it has just one neighbour (with $j$-invariant $a$) with just one dual pair between them. So, 
            $$P(0) = P(0, a) = \dfrac{2}{\frac{p-1}{2}}$$
        Similarly, if a node has $j$-invariant $1728$, then it has one neighbour with $j$-invariant $b \neq 1728$ and also a self-loop, and so
            $$P(1728) = P(1728, 1728) + P(1728, b) = \dfrac{1 + 2}{\frac{p-1}{2}} = \dfrac{3}{\frac{p-1}{2}}$$
        Finally, if a node has $j$-invariant $c \neq 0, 1728$, then it has up to three neighbours, and exactly three dual pairs. Irrespective of how these dual pairs are distributed among the neighbours, the final probability adds up to
            $$P(c) = \dfrac{6}{\frac{p-1}{2}}$$
        
        To see that this is an eigenvector of $M$, we will assume that we are currently at each (current, previous) pair with probabilities as described above. We will then show that moving one more step through the graph does not change these probabilities. This is the same as showing that the vector of these probabilities is unchanged when multiplied by the stochastic matrix $M$ associated to the graph, making it an eigenvector of $M$ with eigenvalue $1$.
        
        So, assume, at step $t$, that the probability $P_t(j_1, j_2)$ of being at each (current, previous) pair is $P(j_1, j_2)$, as in the proposed eigenvector. We work case by case to compute $P_{t+1}(j_1, j_2)$ using the following formula:
            $$P_{t+1}(j_1, j_2) = \sum_{j \in N_{j_2}} P_t(j_2, j) \cdot M_{(j_2,j), (j_1,j_2)}$$
        where $N_{j_2}$ is the set of $j$-invariants that are neighbours of $j_2$, and $M_{(j_2,j), (j_1,j_2)}$ is the entry of $M$ in the $(j_2, j)$ column and $(j_1, j_2)$ row. We recall that this entry of $M$ describes the likelihood to going to $(j_1, j_2)$ from $(j_2, j)$. The cases are as follows:
        \begin{enumerate}
            \item $j_1, j_2 \neq 1728, 0$, and there is one dual pair between the nodes, as seen in \cref{fig:case1}.
            
            \begin{figure}[H]
                \centering
                \includegraphics[scale=.8]{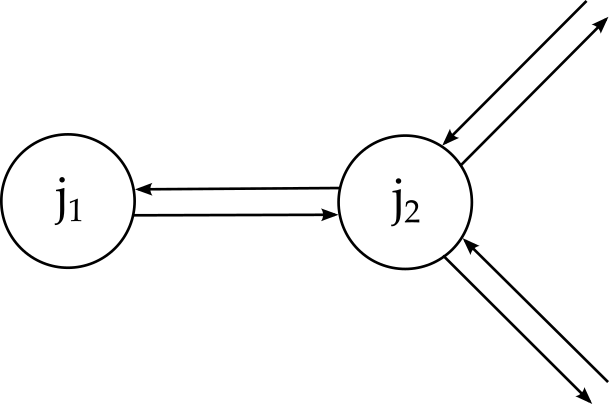}
                \caption{Case 1: $j_1, j_2 \neq 1728, 0$}
                \label{fig:case1}
            \end{figure}
    
            For each arrow pointing at $j_2$ we assume a probability of $\slfrac{2}{\frac{p-1}{2}}$, in accordance with the eigenvector. We see that one arrow comes from $j_1$, while the other two arrows come from elsewhere. If we entered $j_2$ via the arrow from $j_1$, then we cannot backtrack to go to $j_1$. However, if we entered $j_2$ from either of the other arrows, then there is a $0.5$ chance of moving to $j_1$ next. We get the following equation:
                $$P_{t+1}(j_1, j_2) = \dfrac{2}{\frac{p-1}{2}} \cdot 0 + \dfrac{2}{\frac{p-1}{2}} \cdot \frac{1}{2} + \dfrac{2}{\frac{p-1}{2}} \cdot \frac{1}{2} = \dfrac{2}{\frac{p-1}{2}}$$
            which matches the proposed eigenvector. Note that the argument is unchanged when $j_1 = j_2$ and the arrow in question is a self-loop.
            
            \item $j_1, j_2 \neq 1728, 0$, and there are two dual pairs between the nodes, as seen in \cref{fig:case2}
            
            \begin{figure}[H]
                \centering
                \includegraphics[scale=.8]{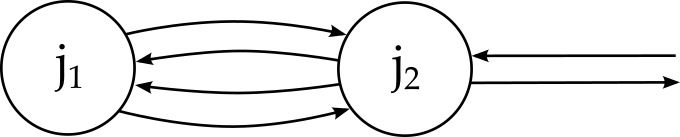}
                \caption{Case 2: $j_1, j_2 \neq 1728, 0$}
                \label{fig:case2}
            \end{figure}
            
            Here, if we entered $j_2$ via an arrow from $j_1$, there is a $0.5$ chance of going back to $j_1$, this time via the other dual pair. However, if we entered $j_2$ from its third arrow, then we are guaranteed to go to $j_1$ next because we cannot backtrack. The equation becomes
                $$P_{t+1}(j_1, j_2) = \dfrac{2}{\frac{p-1}{2}} \cdot \frac{1}{2} + \dfrac{2}{\frac{p-1}{2}} \cdot \frac{1}{2} + \dfrac{2}{\frac{p-1}{2}} \cdot 1 = \dfrac{4}{\frac{p-1}{2}}$$
            which again matches the proposed eigenvector. Once again, this also works when $j_1 = j_2$.
            
            \item $j_1 = 0$ and $j_2 \neq 0, 1728$. By \cref{lemma:j=0}, this looks like \cref{fig:case3}:
            
            \begin{figure}[H]
                \centering
                \includegraphics[scale=.8]{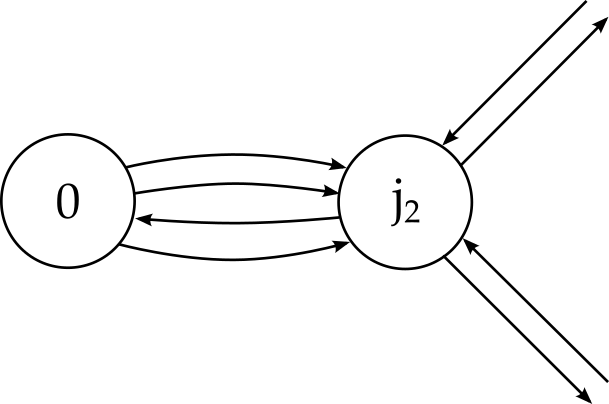}
                \caption{Case 3: $0$ and $ j_2 \neq 1728, 0$}
                \label{fig:case3}
            \end{figure}
            
            Here, if we entered $j_2$ from any of the arrows from $0$ to $j_2$, then we cannot backtrack to $0$, because the sole arrow going backwards is dual to all three incoming arrows. on the other hand, if we entered $j_2$ from elsewhere, there is a $0.5$ chance of advancing to $0$. 
                $$P_{t+1}(0, j_2) = \dfrac{2}{\frac{p-1}{2}} \cdot 0 + \dfrac{2}{\frac{p-1}{2}} \cdot \frac{1}{2} + \dfrac{2}{\frac{p-1}{2}} \cdot \frac{1}{2} = \dfrac{2}{\frac{p-1}{2}}$$
            
            \item $j_1 \neq 0, 1728$ and $j_2 = 0$. By \cref{lemma:j=0}, this looks like \cref{fig:case4}:
            
            \begin{figure}[H]
                \centering
                \includegraphics[scale=.8]{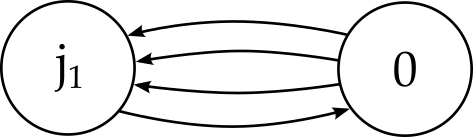}
                \caption{Case 4: $j_1\neq 1728, 0$ and $0$}
                \label{fig:case4}
            \end{figure}
            
            There is only one way of entering $0$, and only one place we can get to from $0$. So,
                $$P_{t+1}(j_1, 0) = \dfrac{2}{\frac{p-1}{2}} \cdot 1 = \dfrac{2}{\frac{p-1}{2}}$$
                
            \pagebreak
            
            \item $j_1 = 1728$ and $j_2 \neq 0, 1728$. By \cref{lemma:j=1728}, this looks like \cref{fig:case5}:
            
            \begin{figure}[H]
                \centering
                \includegraphics[scale=.8]{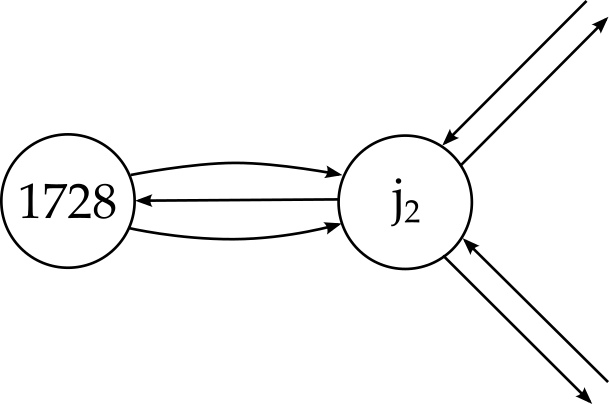}
                \caption[scale=.8]{Case 5: $1728$ and $j_2 \neq 1728,0$}
                \label{fig:case5}
            \end{figure}
            
            If we entered $j_2$ from either of the arrows from $1728$, we cannot backtrack. However, if we entered $j_2$ from elsewhere, there is a $0.5$ chance of advancing to $1728$. 
                $$P_{t+1}(1728, j_2) = \dfrac{2}{\frac{p-1}{2}} \cdot 0 + \dfrac{2}{\frac{p-1}{2}} \cdot \frac{1}{2} + \dfrac{2}{\frac{p-1}{2}} \cdot \frac{1}{2} = \dfrac{2}{\frac{p-1}{2}}$$
            
            \item $j_1 \neq 0, 1728$ and $j_2 = 1728$. By \cref{lemma:j=1728}, this looks like \cref{fig:case6}:
            
            \begin{figure}[H]
                \centering
                \includegraphics[scale=.8]{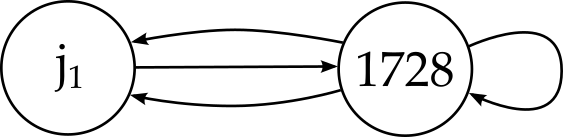}
                \caption{Case 6: $j_1\neq 1728, 0$ and $1728$}
                \label{fig:case6}
            \end{figure}
            
            If we entered $1728$ from $j_1$, then there is a $0.5$ chance of going back to $j_1$ via the other arrow pointing back. However, if we entered $1728$ from $1728$, then we are guaranteed to advance to $j_1$.
                $$P_{t+1}(j_1, 1728) = \dfrac{2}{\frac{p-1}{2}} \cdot \frac{1}{2} + \dfrac{1}{\frac{p-1}{2}} \cdot 1 = \dfrac{2}{\frac{p-1}{2}}$$ 
            \item $j_1 = 1728$ and $j_2 = 0$. Combining both \cref{lemma:j=0} and \cref{lemma:j=1728}, this looks like \cref{fig:case7}:
            
            \begin{figure}[H]
                \centering
                \includegraphics[scale=.8]{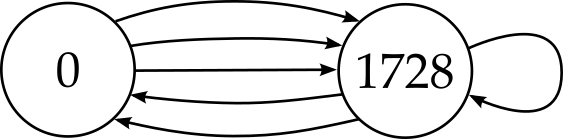}
                \caption{Case 7: $0$ and $1728$}
                \label{fig:case7}
            \end{figure}
            
            There is only one way to get to $0$, and only one way out of $0$. So
                $$P_{t+1}(1728, 0) = \dfrac{2}{\frac{p-1}{2}} \cdot 1 = \dfrac{2}{\frac{p-1}{2}}$$
                
            \item $j_1 = 0$ and $j_2 = 1728$. Combining \cref{lemma:j=0} and \cref{lemma:j=1728}, we get \cref{fig:case8}:
            
            \begin{figure}[H]
                \centering
                \includegraphics[scale=.8]{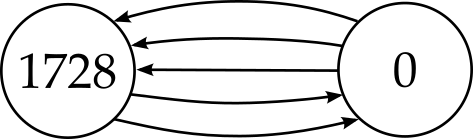}
                \caption{Case 8: $1728$ and $0$}
                \label{fig:case8}
            \end{figure}
            
            If we entered $1728$ from $0$, we have a $0.5$ chance of leaving to $0$. However, if we entered $1728$ from $1728$, we are guaranteed to move to $0$ next.
                $$P_{t+1}(0, 1728) = \dfrac{2}{\frac{p-1}{2}} \cdot \frac{1}{2} + \dfrac{1}{\frac{p-1}{2}} \cdot 1 = \dfrac{2}{\frac{p-1}{2}}$$
        \end{enumerate}
        
        We point out a small but important difference in some of the above casework. In case 3, we saw that if we go from $0$ to $j_2$, we cannot go back to $0$ next, because the sole arrow back is dual to all three arrows from $0$ to $j_1$. But in case 4, we saw that after entering $0$ from $j_1$, we were able to go back to $j_1$, even though the three arrows to $j_1$ are all dual to the one we just came to $0$ from. This is because in the CGL hash function, we disallow backtracking not based on duals, but based on kernels.
        
        In case 3, each isogeny $\phi_i$ from $0$ to $j_2$ is such that $\ker{\hat{\phi}_1} = \ker{\hat{\phi}_2} = \ker{\hat{\phi}_3}$. This is why we cannot backtrack. On the other hand, in case 4, the isogeny $\phi$ from $j_1$ to $0$ has three duals, all with different kernels that are permuted transitively by the automorphism group of $j=0$. Since the kernels are different, we are allowed to backtrack. Similar issues come up in cases 5,6 and in cases 7,8, but can be explained in the same way.
        
        This proves that the proposed values do in fact form an eigenvector of $M$ with eigenvalue $1$, completing the proof.
    \end{proof}
    
    
\section{Conclusions}\label{Sec:conclusions}

    \subsection{Probability of Collisions}
    In this subsection, we use the probability distributions to describe the collision resistance of the CGL hash function. \Cref{Sec:future} contains several interesting directions for future work.
    
    Now that we have the probability distributions for the hash values of every supersingular isogeny graph $G_2(K)$, we can find out how likely it is for two different bitstrings to have a collision. Given a node with $j$-invariant $j$, the probability of two randomly chosen, sufficiently long bitstrings having hash value $j$ is approximately $P(j)^2$. Therefore, the probability of any collision occurring is
        $$\sum_{j \in \F_{p^2} \text{ supersingular}} P(j)^2$$
    In a hash function where all hash values are evenly distributed, so that if $n$ is the number of possible hash values, and each is attained with probability $\frac{1}{n}$, we would expect the probability of a collision to be 
        $$n \cdot \lrp{\frac{1}{n}}^2 = \frac{1}{n}$$
    This is exactly what happens in the CGL hash function when $p \equiv 1 \mod 12$, so that by \cref{Thm:main}, all nodes are evenly distributed. Things are more interesting when $p \not \equiv 1 \mod 12$ so that at least one of $j=0$ and $j=1728$ is supersingular.
    
    For example, when $p \equiv 5 \mod 12$, we see that there are $\frac{p-5}{12}$ nodes with $j \neq 0, 1728$ that are supersingular. Additionally, $j = 0$ is supersingular. So, the probability of a collision is
        \begin{align*}
            \frac{p - 5}{12} \cdot \lrp{\dfrac{6}{\frac{p-1}{2}}}^2 + \lrp{\dfrac{2}{\frac{p-1}{2}}}^2 &= \frac{p-5}{12} \cdot \frac{36 \cdot 4}{(p-1)^2} + \frac{4 \cdot 4}{(p-1)^2} \\
            &= \frac{12p - 44}{(p-1)^2}
        \end{align*}
    Similar calculations reveal that when $p \equiv 7 \mod 12$, the probability of a collision is
        \begin{align*}
            \frac{p - 7}{12} \cdot \lrp{\dfrac{6}{\frac{p-1}{2}}}^2+\lrp{\dfrac{3}{\frac{p-1}{2}}}^2 &= \frac{p-7}{12}\cdot\frac{36\cdot4}{(p-1)^2}+\frac{9\cdot4}{(p-1)^2}\\
            &=\frac{12p-48}{(p-1)^2}
        \end{align*}
    and when $p \equiv 11 \mod 12$, the probability of a collision is
        \begin{align*}
            \frac{p-11}{12}\cdot\lrp{\dfrac{6}{\frac{p-1}{2}}}^2+\lrp{\dfrac{2}{\frac{p-1}{2}}}^2+\lrp{\dfrac{3}{\frac{p-1}{2}}}^2 &=\frac{p-11}{12}\cdot\frac{36\cdot4}{(p-1)^2}+\frac{4 \cdot 4}{(p-1)^2}+\frac{9\cdot4}{(p-1)^2}\\
            &=\frac{12p - 80}{(p-1)^2}
        \end{align*}

    \subsection{Comparing Collision Rates in the Actual and Ideal Cases}
    In an ideal hash function, all hash values would be equally distributed. The above probabilities show that this is not always the case in the CGL hash function. So, we can compare the probability of a collision in the actual case to that in the ideal case, to find how much more likely it is for there to be a collision in the actual case than in the ideal case. To get an idea of the size of this "error", we can compare that value to the likelihood of a cosmic ray error, which is a known source of error in all computing. According to \cite[Ch.~7]{Holden}, if the error is less than the likelihood of a cosmic ray error, we can safely say that the error is negligible. 
    
    When $p \equiv 5 \mod 12$, the number of nodes in the graph is $\frac{p-5}{12} + 1 = \frac{p+7}{12}$. So, if the hash function were evenly distributed, the probability of a collision would be $\frac{12}{p+7}$.
    
\begin{figure}[H]
            \centering
            \begin{tikzpicture}
                \begin{axis}[
                    xmin=0,
                    xmax=100,
                    ymin=-0.5,
                    ymax=0.5,
                    axis lines=middle,
                    xlabel=$p$,
                    ylabel=$e$,
                    disabledatascaling,
                    trig format plots=rad,]
                       
                        \addplot[domain=5:100, samples = 300, red, thick]{(64*x - 320)/((x^2 - 2*x + 1)*(x+7))};
                        \addlegendentry{$p \equiv 5 \mod 12$}
                        \addplot[domain=7:100, samples = 300, blue, thick]{(36*x - 252)/((x^2 - 2*x + 1)*(x+5))};
                        \addlegendentry{$p \equiv 7 \mod 12$}
                        \addplot[domain=10.63:100, samples = 300, green, thick]{(100*x - 1052)/((x^2 - 2*x + 1)*(x+13))};
                        \addlegendentry{$p \equiv 11 \mod 12$}
                    \end{axis}
            \end{tikzpicture}
            \caption{\centering difference in probability of a collision in the actual case and the ideal case}
            \label{fig:error}
        \end{figure}
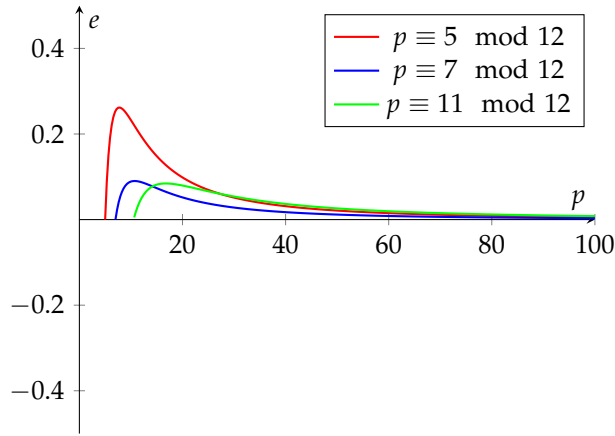

    \pagebreak
    We compute the difference in probabilities:
        $$\frac{12p - 44}{(p-1)^2} - \frac{12}{p+7} = \frac{(12p - 44)(p+7) - 12(p-1)^2}{(p-1)^2(p+7)} = \frac{64p - 320}{p^3 + 5p^2 - 13p + 7}$$
    Similarly, we can compute this difference when $p \equiv 7 \mod 12$, where the number of nodes is $\frac{p+5}{12}$.
        $$\frac{12p - 48}{(p-1)^2} - \frac{12}{p+5} = \frac{36p - 252}{p^3 + 3p^2 - 9p + 5}$$
    and once again when $p \equiv 11 \mod 12$, where the number of nodes is $\frac{p+13}{12}$.
        $$\frac{12p - 80}{(p-1)^2} - \frac{12}{p+13} = \frac{100p - 1052}{p^3 + 11p^2 - 26p + 13}$$
    Even though we only care about the above values when $p \equiv 5, 7, 11 \mod 12$ respectively, we can plot the functions over the real numbers to get an idea of their long term behaviour. As we can see in \cref{fig:error}, the error tends to $0$ as $p$ is increased in all three cases.
        
        The above error values are only accurate when the hashed bitstring is sufficiently long. So, we assume a standard file size of $1$MB, or $8,000,000$ bits. According to \cite[Sec.~ 4.2.1]{CGL}, the hash function runs at a speed of $13.1$Kbps when the prime used is $256$ bits long. This means that it takes $610.687$s to hash $1$MB of data. According to \cite[Sec.~III.B]{Slayman}, the mean time between consecutive cosmic ray errors ranges anywhere between $1$ and $500$ years. $1$ year contains $31536000$ seconds, and so each year, one could hash $51640.202$MB of data. This means that the expected number of cosmic ray errors per MB of data ranges between $1.94 \cdot 10^{-5}$ and $3.87 \cdot 10^{-8}$.
        
        Since the cosmic ray error was computed based on a $256$ bit prime, we can assume that $p \approx 2^{255}$. Plugging this into our error functions yields
            \begin{equation*}
                \text{Error} = \begin{cases}
                    1.91 \cdot 10^{-152}, & p \equiv 5 \mod 12 \\
                    1.07 \cdot 10^{-152}, & p \equiv 7 \mod 12 \\
                    2.98 \cdot 10^{-152}, & p \equiv 11 \mod 12
                \end{cases}
            \end{equation*}
        All three values are far below the likelihood of a cosmic ray error, and so we can say that from a practical standpoint, the theoretical imperfections of the CGL hash function are negligible.
    
    \subsection{Future Work}\label{Sec:future}
        This work could be continued with an investigation into the minimum bitstring length required to reach probability distributions within $e$ of the expected probabilities, for some predetermined error bound $e$. It may be useful to find a relationship between the length of the bitstring and the error of the probability distributions. This way, we could come up with approximations for probability distributions based on shorter bitstrings.
        
        Another direction could involve taking the stochastic matrix for our graph and using it as the adjacency matrix for a new graph. This new graph could yield interesting results or insights into the original graph. 
        
        Finally, we could also repeat the work done in this paper for higher degree isogenies. It might be possible to generalize our formula for probability distributions based on $l$, the degree of the isogenies.
\pagebreak   

\printbibliography[heading=bibintoc,title={References}]

@book{Silverman,
  author =       "Joseph H. Silverman",
  title =        {Arithmetic of Elliptic Curves.},
  year =         "1997",
  DOI ="https://doi.org/10.1017/cbo9781139174879.008",
  publisher="Springer-Verlag New York"
}

@book{Galbraith,
    author="Steven D. Galbraith.",
    title={Mathematics of Public Key Cryptography},
    year="2012",
    DOI="https://doi.org/10.1017/CBO9781139012843",
    publisher={Cambridge University Press, USA.}
}

@thesis{Kohel,
    author={David Kohel},
    title={Endomorphism rings of elliptic curves over finite fields},
    school="University of California at Berkeley",
    year={1996},
    url={http://iml.univ-mrs.fr/~kohel/pub/thesis.pdf}
}

@article{CGL,
    author="D.X. Charles and E.Z. Goren and K.E. Lauter",
    title={Cryptographic Hash Functions from Expander Graphs},
    year={2009},
    DOI="https://doi.org/10.1007/s00145-007-9002-x",
    pages={93-113},
    journal="Journal of Cryptology",
    volume ="22"
}

@article{Adj,
    author="Gora Adj and Omran Ahmadi and Alfred Menezes",
    title="On Isogeny Graphs of Supersingular Elliptic Curves over Finite Fields.",
    journal="Finite Fields and Their Applications",
    volume="55",
    year="2019",
    DOI="https://doi.org/10.1016/j.ffa.2018.10.002",
    pages={268-283}
}

@book{Lang,
    author="Serge Lang",
    title= "Algebra",
    year ={2002},
    DOI="https://doi.org/10.1007/978-1-4613-0041-0",
    publisher="Springer-Verlag New York"
}

@book{hac,
    title="Handbook of Applied Cryptography",
    author="Alfred J. Menezes and Paul C. van Oorschot and Scott A. Vanstone",
    year={1996},
    publisher="CRC Press",
    ISBN="0-8493-8523-7"
}

@book{Lay,
    author="David C. Lay",
    title="Linear Algebra and its Applications",
    year="2016",
    edition="5",
    ISBN="978-0-321-98238-4",
    publisher="Pearson Education, Inc."
}

@article{Munuera,
    author="C. Munuera and J. Tena",
    title="An algorithm to compute the number of points on elliptic curves of $j$-invariant 0 or 1728 over a finite field",
    journal ={Rendiconti Del Circolo Matematico Di Palermo},
    series="II",
    volume="42",
    year={1993},
    pages={106-116}
}

@book{Holden,
    author="Joshua Holden",
    title="The Mathematics of Secrets",
    year={2017},
    publisher="Princeton University Press",
    ISBN="978-0-691-14175-6",
}

@article{Slayman,
    author="C. Slayman",
    title="Cache and Memory Error Detection, Correction, and Reduction Techniques for Terrestrial Servers and Workstations",
    journal={IEEE Transactions on Device and Materials Reliability},
    volume="5",
    year={2005},
    pages={397-404}
}

\end{document}